\documentclass[fleqn,usenatbib]{mnras}

\usepackage{newtxtext,newtxmath}

\usepackage[T1]{fontenc}

\DeclareRobustCommand{\VAN}[3]{#2}
\let\VANthebibliography\thebibliography
\def\thebibliography{\DeclareRobustCommand{\VAN}[3]{##3}\VANthebibliography}

\usepackage{graphicx}	
\usepackage{amsmath}	

\title[Cosmology with WL peak redshifts]{The FLAMINGO project: cosmology with the redshift dependence of weak gravitational lensing peaks}

\author[J. C. Broxterman et al.]{
Jeger C. Broxterman,$^{1,2}$\thanks{E-mail: broxterman@lorentz.leidenuniv.nl}
Matthieu Schaller,$^{1,2}$ 
Henk Hoekstra,$^{2}$ 
Joop Schaye,$^{2}$ 
Robert J. McGibbon,$^{2}$ \newauthor
Victor J. Forouhar Moreno$^{2}$,
Roi Kugel$^{2}$
and Willem Elbers$^{3}$ 
\\
$^{1}$Lorentz Institute for Theoretical Physics, Leiden University, PO Box 9506, NL-2300 RA Leiden, the Netherlands
\\
$^{2}$Leiden Observatory, Leiden University, PO Box 9513, NL-2300 RA Leiden, The Netherlands \\
$^{3}$Institute for Computational Cosmology, Department of Physics, University of Durham, South Road, Durham, DH1 3LE, UK
}

\date{Accepted XXX. Received YYY; in original form ZZZ}

\pubyear{2024}

\begin{document}
\label{firstpage}
\pagerange{\pageref{firstpage}--\pageref{lastpage}}
\maketitle

\begin{abstract}
Weak gravitational lensing (WL) convergence peaks contain valuable cosmological information in the regime of non-linear collapse. Using the FLAMINGO suite of cosmological hydrodynamical simulations, we study the physical origin and redshift distributions of the objects generating WL peaks selected from a WL convergence map mimicking a \textit{Euclid} signal. We match peaks to individual haloes and show that the high signal-to-noise ratio (SNR~>~5) WL peaks measured by Stage IV WL surveys primarily trace $M_{\mathrm{200c}} > 10^{14}~\mathrm{M_\odot}$ haloes. We find that the WL peak sample can compete with the purity and completeness of state-of-the-art X-ray and  Sunyaev-Zel’dovich cluster abundance inferences. By comparing the distributions predicted by simulation variations that have been calibrated to the observed gas fractions of local clusters and the present-day galaxy stellar mass function, or shifted versions of these, we illustrate that the shape of the redshift distribution of SNR~>~5 peaks is insensitive to baryonic physics while it does change with cosmology. The difference highlights the potential of using WL peaks to constrain cosmology. As the WL convergence and redshift number densities of WL peaks scale differently with cosmology and baryonic feedback, WL peak statistics can simultaneously calibrate baryonic feedback and constrain cosmology.
\end{abstract}




\section{Introduction}
The cornerstone of modern cosmology is the $\Lambda$CDM model, consisting primarily of dark energy ($\Lambda$) and cold dark matter (CDM). The model can predict a wide range of cosmological observables, including the cosmic microwave background \citep[CMB, e.g.][]{Planck2020}, baryonic acoustic oscillations \citep[BAO, e.g.][]{DESI2024}, and measurements of the luminosity distance ($D_L$) from supernovae \citep[SNe, e.g.][]{Riess2021}. Other probes, such as redshift-space distortions (RSDs) \citep[e.g.][]{Scoccimarro2004}, galaxy clustering \citep[e.g.][]{Cacciato2009}, CMB lensing \citep[e.g.][]{Planck2020_CMBlensing}, and cosmic shear measure the properties of the large-scale structure (LSS) over a range of scales and with cosmic time. Cosmic shear is the distortion of galaxy shapes by weak gravitational lensing (WL) by the large-scale structure, and it is a powerful tool to map the matter distribution in the universe \citep[for a review, see e.g.][]{kilbinger2015cosmology}. Many of the LSS probes report values of the $S_\mathrm{8}$ parameter that are in mild tension with CMB predictions. This parameter is well constrained by WL inferences and measures a combination of the matter density and the amplitude of density fluctuations. The $S_\mathrm{8}$ tension manifests itself between CMB and cosmic shear \citep[e.g.][]{Hildebrandt2017,asgari2021kids,Secco2022,Abbott2023}, galaxy clustering \citep[e.g.][]{Troster2020,Philcox2022}, RSDs \citep[e.g.][]{Benisty2021}, and thermal Sunyaev-Zel’dovich (tSZ) measurements \citep[e.g.][]{McCarthy2014,McCharty2023}. 

An additional statistic that is used to constrain cosmology is cluster counts \citep[e.g.][]{Wang1998,Reiprich2002,Mana2013,Fumagalli2024}. Clusters are the largest virialized structures in the universe, and their number density probes the high-mass end of the halo mass function (HMF), which is sensitive to the amount of matter ($\Omega_\mathrm{m}$), its clumpiness ($\sigma_8$), and the evolution of dark energy ($w_0, w_a$) \citep[e.g.][]{Courtin2011,Bocquet2016}.  For a review, see \citet{Allen2011}.

Clusters may be selected through different observables (e.g. optical richness, SZ or X-ray), but selection effects and model assumptions typically limit cluster cosmology inferences \citep[e.g.][]{Nagai2007,Kugel2024syserrors}. Whereas a Compton-y selected sample can provide cleaner selection effects, and more accurate mass estimates can be obtained by calibrating mass proxies using lensing measurements, the approach still suffers from not being able to definitively determine the completeness of the sample as clusters may be missing from the sample due to variations in their temperature and gas content \citep[e.g.][]{Kugel2024}. Additionally, current cluster inferences require several model assumptions, such as the choice of HMF, the assumed functional form for observable-mass scaling relations and the associated scatter, which may lead to biased results for next-generation measurements that aim to do precision cosmology at the sub-per cent level \citep[][]{Bocquet2016,Kugel2024syserrors}. 

In this paper, we instead explore a different approach that aims to extract cosmological information by also probing the high-mass end of the HMF as a function of redshift, but that suffers less from selection effects and model assumptions as the selection is carried out using a WL convergence ($\kappa$) map only. By selecting peaks from a WL map, we expect to extract the same (and more) information probed by current cluster analyses, as we will show that the peaks correspond to virtually all the most massive objects in the universe leading to a pure and complete sample. Crucially, the WL selection of these objects is not plagued by the uncertainties originating from assumptions regarding the dynamical state of the galaxy clusters (e.g. relaxedness or hydrostatic equilibrium) nor by scaling relation calibration problems.

This paper focuses on the local maxima in WL convergence maps, commonly referred to as WL peaks \citep[see e.g.][]{dietrich2010cosmology,Kacprzak2016,Martinet2018,davies2021constraining,davies2022cosmological,Li2023}. These peaks are a non-Gaussian statistic and correspond to the most massive objects in the universe that have undergone non-linear collapse and, therefore, contain a wealth of cosmological information \citep[see e.g.][]{kratochvil2010probing}. The origin of WL peaks in the context of single, massive objects has been studied using simulations by \citet{Hamana2004}, who found that high signal-to-noise ratio (SNR) peaks originate from single massive objects along the line of sight as opposed to several haloes aligned along the line of sight. These results were confirmed by \citet{yang2011cosmological} and \citet{liu2016origin}, who also found that peaks with SNR~>~3.5 can generally be attributed to a single massive halo. The studies broadly agree, but as there are differences in redshift distributions, noise values, methodology, and SNR cuts, the three studies differ in their exact predictions. We aim to further this understanding by considering the analysis in the context of Stage IV WL surveys, i.e. \textit{Euclid} \citep[][]{EuclidPaper1_2024}{}{}, \textit{Roman} \citep[][]{spergel2015wide}{}{}, and \textit{Rubin} \citep[][]{abell2009lsst}{}{}, and by matching individual peaks to haloes to determine the halo mass regimes corresponding to WL peaks of different WL convergence values.

The WL convergence value of a WL peak depends on the lens mass and redshift (as this impacts the WL kernel) and consequently depends on structure formation as more massive structures form at lower redshift, but also on baryonic feedback as this removes mass from the haloes that act as lenses \citep[see e.g.][]{Stanek2009,Martizzi2014,Velliscig2014,Debackere2022}{}{}. In \citet{Broxterman2024}, we studied how baryonic physics and changes in cosmology impact the counts of WL peaks using the FLAMINGO simulation suite \citep[][]{schaye2023flamingo,kugel2023flamingo}{}{}. There, we showed that the impact of baryonic physics and cosmology on the number density of high-valued WL peaks can be qualitatively understood by considering the impact of feedback processes on the mass of a single halo and the impact of the cosmological parameters (primarily $\Omega_\mathrm{m}$ and $\sigma_\mathrm{8}$) on the HMF. Feedback removes gas from the centres of haloes, causing them to be less overdense. Consequently, the WL effect will be smaller, and the peaks they generate will have lower WL convergence values. Hence, there will be fewer (high-valued) $\kappa$ peaks when haloes experience stronger feedback. Regarding the impact of cosmology, $\Omega_\mathrm{m}$ broadly moves the amplitude of the HMF vertically, and $\sigma_\mathrm{8}$ shifts its exponential cut-off, both causing there to be more/fewer massive haloes in the universe in general, which in turn is also reflected in the number of (high-valued) $\kappa$ peaks. Whereas the impact of realistic baryonic physics is smaller than currently reasonable changes in cosmology, the two signatures are degenerate, complicating a hypothetical cosmological inference. 

In this work, we aim to extend this analysis by including the redshift information of the WL peaks. While the number density of WL peaks in $\kappa$ depends on cosmology and baryonic physics, we expect the formation time of massive objects and corresponding WL peak redshift to depend primarily on cosmology. The formation time of massive haloes should, to first order, not depend on the exact strength of baryonic feedback. Therefore, the redshift distribution of the peaks contains information on when and how much structure in the universe has formed that could be used as an additional probe to constrain cosmology. The redshift dependence of the number density of WL peaks corresponding to these massive objects should be insensitive to baryonic physics. Therefore, we aim to explore the possibility of using the redshift dependence of the number density of WL peaks to constrain cosmology and discriminate between the impact of baryonic feedback and changes in cosmology. 

Recently, \citet{Joachim2024} carried out an inference using WL peaks selected on tomographic WL convergence maps from the KiDS-1000 and DES Y1 data releases. They found that this can greatly improve the results compared with selecting peaks from a single integrated WL convergence map as it can probe the evolution of the growth of structure, already highlighting the possibility of including redshift information. We aim to assign a redshift to every peak and thus expect to include more redshift information than in their study. Also, \citet{Chen2024} and \citet{Chiu2024} carried out a similar analysis as the idea explored in this paper by using HSC Y3 data and by selecting clusters through the peaks of WL aperture mass maps. Their sample only consists of 149 clusters selected through WL, and they thus cannot yet provide meaningful constraints. Nonetheless, it highlights the viability of the approach. For a recent review of peaks selected from aperture mass maps, see \citet{Oguri2024}.

This paper aims to explore in more detail the origin of WL peaks by linking them to individual haloes and leveraging the baryonic insensitivity of their redshift distribution to illustrate their potential to better constrain cosmological parameters. To this end, we use the baryonic physics and cosmology variations of the cosmological hydrodynamical simulation suite FLAMINGO \citep[][]{schaye2023flamingo,kugel2023flamingo}{}{}. We first briefly summarize the relevant WL theory in Section~\ref{sec:theory}. We generate full-sky WL convergence maps mimicking the signal of Stage IV WL surveys, as described in Section~\ref{sec:methods}, where we also introduce the algorithm to match individual peaks to massive haloes each. The results of the halo-peak matching and the evolution of the redshift number density of WL peaks are presented in Section~\ref{sec:results}. The main conclusions are summarized in Section~\ref{sec:sum_and_concl}.

\section{Theory}\label{sec:theory}
As we will construct full-sky WL maps from discretized mass shells spaced linearly in redshift, we first summarize the main equations relevant to WL. For full reviews of the WL formalism, see \citet{bartelmann2001weak,hoekstra2008weak,kilbinger2015cosmology}.  In WL, where the light deflections are small, the magnification matrix ($\boldsymbol{A}$), which describes the linear transformation from the observed angular position ($\boldsymbol{\theta}$) to the unlensed position, is typically decomposed as  
\begin{equation}
A_{ij} = \frac{\partial (\delta \theta_i)}{\delta \theta_j} \approx 
\begin{pmatrix}
1 - \kappa - \gamma_1 & -\gamma_2 \\
-\gamma_2 & 1-\kappa+\gamma_1 
\end{pmatrix},
\end{equation}
where $\kappa$ is the WL convergence and $\gamma = \gamma_1+i \gamma_2$ is the WL shear. When assuming the deflections are small, the deflection angle can be expressed as the gradient of the potential ($\Phi$) that is the solution of the 2D Poisson equation, $\nabla^2\Phi = 4\pi G a^2 \bar{\rho} \delta$, where $G$ is the gravitational constant, $a$ the scale factor, $\bar{\rho}$ the mean density of the universe, and $\delta$ the matter overdensity. By assuming the Born approximation, i.e. evaluating the lensing equations on unperturbed photon paths, the magnification matrix reduces to 
\begin{align}
A_{ij} = \delta_{ij} - \partial_i \partial_j \psi,
\end{align}
with $\psi$ the lensing potential,
\begin{align}
    \psi(\boldsymbol{\theta},\chi) = \frac{2}{c^2} \int_0^\chi \mathrm{d}\chi'\, \frac{\chi - \chi'}{\chi \chi'}\, \Phi(\chi'\boldsymbol{\theta},\chi'),
\end{align}
where $c$ is the speed of light, and we assumed a flat universe such that the comoving angular diameter distance, $f_K(\chi)$, reduces to the comoving distance, $\chi$. By substituting the Poisson equation, the WL convergence can be estimated as
\begin{align}\label{eqn:ana_WL_conv}
    \kappa(\boldsymbol{\theta},\chi) = \frac{3 \Omega_\mathrm{m} H_0^2}{2c^2} \int_0^{\chi_{\mathrm{hor}}} \mathrm{d}\chi \,[1+z(\chi)]\, W(\chi)\, \delta(\chi,\boldsymbol{\theta}),
\end{align}
where $H_0$ is the Hubble constant, $\Omega_\mathrm{m} \equiv \rho_\mathrm{m,0} / \rho_{\mathrm{crit},0}$ is the matter density parameter, $\rho_{\mathrm{crit},0} = 3H_0^2 / (8\pi G) $ is the present-day critical density of the universe, and $z$ is the redshift. $\chi_{\mathrm{hor}}$ is the comoving line of sight horizon to the edge of the galaxy sample. The weak lensing kernel $W$ is given by \citep[][]{Kaiser1992}{}{}
\begin{align}\label{eqn:ana_lensing_kernel}
    W(\chi) = \chi \int_\chi^{\chi_{\mathrm{hor}}} \mathrm{d}\chi'\, n_\textrm{s}(\chi')\, \frac{\chi' - \chi}{\chi'},
\end{align}
where $n_\mathrm{s}(\chi)$ is the comoving source distribution which is normalised ($\int n_\mathrm{s}(\chi) \, \mathrm{d}\chi =1$) and is related to the source redshift distribution as $n_\mathrm{s}(z) \mathrm{d}z = n_\mathrm{s}(\chi) \mathrm{d} \chi$. 

\cite{lu2021impact} have argued that for a Stage III WL peak inference assuming the Born approximation leads to biased results, and a more sophisticated approach such as backward ray tracing should be applied. In this paper, we do not aim to carry out parameter inference but are instead interested in the origin and redshift evolution of a sample of high-valued WL peaks. Although the exact number of WL peaks at fixed WL convergence values may differ slightly depending on the map construction algorithm, we do not expect the origin of these peaks to change, as they should still stem from the same massive structures in the universe. As the Born approximation facilitates a more straightforward and computationally less expensive way to assign a redshift (as will become clear in Section~\ref{sec:peaks}), we adopt it in our analysis. 
Additionally, as we ignore several second-order effects such as lens-lens coupling \citep[][]{Bernardeau1997}, and possible biasing factors such as dilution by cluster member contamination \citep[][]{Applegate2014,Medezinski2018}, the analysis is to some extent idealized. Some of these effects, such as source clustering, are expected to have a minor impact on WL peak statistics \citep[][]{Gatti2024}. For others, such as galaxy intrinsic alignment, the impact, although small, is non-negligible \citep[][]{Zhang2022}. As the impact of these effects is less than 1 per cent on the power spectrum \citep[][]{kilbinger2015cosmology}, we do not expect them to impact the conclusions of this work as we study the relative differences between the different simulation variations, but an analysis using legacy Stage IV WL survey data should consider them in more detail.

\section{Methods}\label{sec:methods}
\subsection{FLAMINGO}
We use the recently developed FLAMINGO cosmological hydrodynamical simulation suite for our analysis. In this section, we only provide a brief summary of the simulation characteristics; for a detailed description of the simulations and calibration, see \citet{schaye2023flamingo} and \citet{kugel2023flamingo}. The simulations were run using the SPHENIX smoothed particle hydrodynamics implementation \citep[][]{borrow2022sphenix}{}{} in SWIFT \citep[][]{schaller2023swift}{}{}. The simulations include an element-by-element based radiative cooling and heating \citep[][]{Ploeckinger2020rates}{}{}, star formation
\citep[][]{Schaye2008stars}{}{}, time-dependent stellar mass loss \citep[][]{Wiersma2009sml}{}{}, and massive neutrinos \citep[][]{elbers2021optimal}{}{}. Supernova and stellar feedback is kinetic and conserves energy and linear and angular momentum \citep[][]{DallaVecchia2008SNfeedback,Chaikin2023SNfeedback}{}{}. The AGN feedback is either thermal \citep[][]{Booth2009AGN}{}{} or kinetic jet-like, where particles are kicked along the black hole (BH) spin axis \citep[][]{Husko2022jets}{}{}. The simulations are calibrated using machine learning to the galaxy stellar mass function at $z=0$ and the gas fraction of low-$z$ clusters and groups of clusters, or systematic shifts in these observables, while taking into account observational uncertainties and biases \citep[][]{kugel2023flamingo}{}{}.

In Table~\ref{tab:simulations}, we list the simulation parameters directly relevant to the interpretation of our results. The first column lists the simulation's identifier, which is set by the box size in comoving Gpc (cGpc) and log$_{10}$ baryonic particle mass. For example, the flagship simulation, L2p8$\_$m9, has 5040$^3$ baryonic and dark matter (DM) particles and 2800$^3$ massive neutrino particles in a 2.8~cGpc box and an initial mean gas particle mass of $1.07\times10^9~\mathrm{M_\odot}$. All 7 cosmology and 8 baryonic feedback variations were carried out in 1~cGpc boxes with an equal number of particles. The fiducial model in this box is L1$\_$m9. L1$\_$m8 and L1$\_$m10 are variations with the same cosmology and calibrated to the same observational data but with 8 times more and fewer particles, respectively. We will use these to quantify the numerical convergence with resolution. All cosmology variations have an accompanying CDM + neutrino run identified by the post-fix $\_$DMO.

The eight variations with different baryonic feedback prescriptions were each calibrated to a (shifted) set of observables, as indicated by the fourth and fifth columns in Table~\ref{tab:simulations}. These variations vary the cluster gas fractions ($\Delta f_{\mathrm{gas}}$) and/or the galaxy stellar mass function ($\Delta M_*$) by a set amount of observational standard deviations ($\sigma$). The baryonic variations are identified by their shifts in observables, as indicated in the first column of Table~\ref{tab:simulations}.

All baryonic variations assume the DES Y3 ‘3 $\times$ 2pt + All Ext.’ best-fitting $\Lambda$CDM cosmology 
from \citet{abbott2022dark}, see also Table~\ref{tab:simulations}. Additional cosmology variations were run using the \citet{Planck2020} best-fitting $\Lambda$CDM model assuming $\sum m_\nu c^2$ = 0.06~eV (‘Planck’), and variations with heavier neutrinos, $\sum m_\nu c^2$ = 0.24~eV  (`PlanckNu0p24Var', `PlanckNu0p24Fix') and $\sum m_\nu c^2$ = 0.48~eV (`PlanckNu0p48Fix'). The PlanckNu0p24Var and PlanckNu0p24Fix differ in their treatment of the other cosmological parameters while changing the neutrino mass. We include two variations with decaying cold dark matter (DCDM) from \citet{Elbers2024}. In these simulations, dark matter decays to dark radiation as specified by the decay rate $\Gamma$ in units $H_0/h$, as indicated by the final column in Table~\ref{tab:simulations}. Finally, we use a simulation with a ‘lensing cosmology’ with a lower value of $\sigma_{8}$ (`LS8') from \citet{amon2023consistent} that is consistent with BOSS DR12 galaxy clustering \citep[][]{Reid2016}, and galaxy-galaxy lensing from KiDS-1000 \citep[][]{asgari2021kids}, DES Y3 \citep[][]{abbott2022dark}, and HSC Y1 \citep[][]{Aihara2018}.

\begin{table*}
\caption{Simulation variations. From left to right, the columns list the simulation identifier; the box size, $L$; the number of baryonic and dark matter particles, $N_\mathrm{p}$; the initial mean baryonic particle mass, $m_{\mathrm{b}}$; the shift in the number of standard deviations ($\sigma$) applied to the observational data used in the calibration for the $z=0$ galaxy stellar mass function ($\Delta M_*$) and the gas fraction of low-redshift clusters ($\Delta f_{\mathrm{gas}}$); the mode of AGN feedback; the dimensionless Hubble constant, $h$; the present-day total matter density parameter, $\Omega_{\mathrm{m}}$; the present-day baryonic matter density parameter, $\Omega_{\mathrm{b}}$; the sum of neutrino species particle masses, $\sum m_\nu c^2$; the amplitude of the initial power spectrum parametrized as the rms mass density fluctuation in spheres with radius 8~Mpc/$h$ extrapolated to $z=0$ using linear theory, $\sigma_8$ and, if applicable, the dark matter decay rate, $\Gamma$.}
\label{tab:simulations}
\begin{tabular}{llllrrlllllll}
\hline
Identifier      & L & $N_\mathrm{p}$ & $m_\mathrm{b}$ & $\Delta M_*$ & $\Delta f_{\mathrm{gas}}$ & AGN & $h$   & $\Omega_{\mathrm{m}}$ & $\Omega_{\mathrm{b}}$ &  $\sum m_\nu c^2$  & $\sigma_8$ & $\Gamma$ \\
               & cGpc & & $\times 10^9~\mathrm{M_\odot}$ & $\sigma$ & $\sigma$ & &  &            &   & eV &   &  $H_0/h$  \\
\hline
L2p8$\_$m9     & 2.8 & 5040$^3$ & 1.07  & 0 & 0 & thermal &  0.681 & 0.306 & 0.0486 & 0.06 &  0.807 & $-$ \\
L1$\_$m9       & 1 & 1800$^3$ & 1.07 & 0 & 0 & thermal & 0.681 & 0.306 & 0.0486 & 0.06 &  0.807 & $-$ \\
L1$\_$m8       & 1 & 3600$^3$ & 0.134 & 0 & 0 & thermal  & 0.681 & 0.306 & 0.0486 & 0.06 &  0.807 & $-$ \\
L1$\_$m10      & 1 & 900$^3$ & 8.56  & 0 & 0 & thermal & 0.681 & 0.306 & 0.0486 & 0.06 &  0.807 & $-$ \\
fgas$+2\sigma$ & 1 & 1800$^3$ & 1.07  & 0 & $+2$ & thermal & 0.681 & 0.306 & 0.0486 & 0.06 &  0.807 & $-$ \\
fgas$-2\sigma$ & 1 & 1800$^3$ & 1.07  & 0 & $-2$ & thermal & 0.681 & 0.306 & 0.0486 & 0.06 &  0.807 & $-$ \\
fgas$-4\sigma$ & 1 & 1800$^3$ & 1.07 & 0 & $-4$ & thermal & 0.681 & 0.306 & 0.0486 & 0.06 &  0.807 & $-$ \\
fgas$-8\sigma$ & 1 & 1800$^3$ & 1.07  & 0 & $-8$ & thermal & 0.681 & 0.306 & 0.0486 & 0.06 &  0.807 & $-$ \\
M*$-\sigma$    & 1 & 1800$^3$ & 1.07  & $-1$ & 0 & thermal & 0.681 & 0.306 & 0.0486 & 0.06 &  0.807 & $-$ \\
M*$-\sigma\_$fgas$-4\sigma$     & 1 & 1800$^3$ & 1.07  & $-1$ & $-4$ & thermal & 0.681 & 0.306 & 0.0486 & 0.06 &  0.807 & $-$ \\
Jet            & 1 & 1800$^3$ & 1.07  & 0 & 0 & jet & 0.681 & 0.306 & 0.0486 & 0.06  & 0.807 & $-$ \\
Jet$\_$fgas$-4\sigma$      & 1 & 1800$^3$ & 1.07  & 0 & $-4$ & jet & 0.681 & 0.306 & 0.0486 & 0.06 & 0.807 & $-$ \\
Planck          & 1 & 1800$^3$ & 1.07  & 0 & 0 & thermal & 0.673 & 0.316 & 0.0494 & 0.06 &  0.812  & $-$ \\
PlanckNu0p24Var & 1 & 1800$^3$ & 1.06  & 0 & 0 & thermal & 0.662 & 0.328 & 0.0510 & 0.24 &  0.772  & $-$  \\
PlanckNu0p24Fix & 1 & 1800$^3$ & 1.07  & 0 & 0 & thermal & 0.673 & 0.316 & 0.0494 & 0.24 &  0.769 & $-$  \\
PlanckNu0p48Fix & 1 & 1800$^3$ & 1.07  & 0 & 0 & thermal & 0.673 & 0.316 & 0.0494 & 0.48 &  0.709 & $-$  \\
PlanckDCDM12    & 1 & 1800$^3$ & 1.07  & 0 & 0 & thermal & 0.673 & 0.274 & 0.0494 & 0.06 &   0.794 & 0.12  \\
PlanckDCDM24    & 1 & 1800$^3$ & 1.07  & 0 & 0 & thermal & 0.673 & 0.239 & 0.0494 & 0.06 &   0.777 & 0.24  \\
LS8             & 1 & 1800$^3$ & 1.07  & 0 & 0 & thermal & 0.682 & 0.305 & 0.0473 & 0.06 &  0.760 & $-$  \\
\hline
\end{tabular}
\end{table*}

\subsubsection{Mass maps}\label{sec:mass_maps}
We use the FLAMINGO mass maps to construct our WL convergence maps. These are a set of concentric spherical HEALPix \citep[][]{gorski2005healpix}{}{} surfaces on which the particles that contribute to an observer's past light-cone are projected \citep[see Appendix A2 of][]{schaye2023flamingo}{}{}. The L1 and L2p8 simulations have 2 and 8 virtual observers, respectively. The light-cone consists of 60 shells of equal thickness (in redshift) between $z=0$ and $3$ ($\Delta z = 0.05$). These maps are the input for constructing the WL convergence maps. To allow the use of all functionality of \textsc{healpy} \citep[][]{Zonca2019Healpy}{}{}, we downsample the FLAMINGO maps to $N_{\mathrm{side}} = 8192$, corresponding to an angular resolution of $\approx$ 25.8~arcsec. We then transform the total matter shells at each shell $i$ into overdensity shells, $\delta_i(\boldsymbol{\theta})$, using 
\begin{align} \label{eqn:overdensity}
\delta_i(\boldsymbol{\theta}) = \frac{\Sigma_i(\boldsymbol{\theta}) - \overline{\Sigma_i}}{\overline{\Sigma_i}},
\end{align} 
where $\Sigma_i(\boldsymbol{\theta})$ is the surface matter density at shell $i$ at position $\boldsymbol{\theta}$ and $\overline{\Sigma_i}$ is the mean surface matter density, which we evaluate directly from the shell. Because of the discrete redshifts of the FLAMINGO mass maps, we transform the analytic expressions to estimate the WL convergence (equations~\ref{eqn:ana_WL_conv}~\&~\ref{eqn:ana_lensing_kernel}) to discrete sums given by 
 \citep[e.g.][]{mccarthy2018bahamas}{}{}
\begin{align}\label{eqn:kappa_born_discreet}
   \kappa(\boldsymbol{\theta}) =  \frac{3 \Omega_{\mathrm{m}} H_0^2}{2c^2} \sum_{i=1}^{N}\, [1+z(\chi_i)]\, W(\chi_i)\, \delta_i(\boldsymbol{\theta})\, \Delta \chi_i,
\end{align}
and 
\begin{align}\label{eqn:lensing_kernel_discreet}
    W(\chi_i) = \chi_i \sum_{j=i}^N\, \Delta \chi_j \, n_\mathrm{s}(\chi_j)\, \frac{\chi_j - \chi}{\chi_j},
\end{align}
respectively. In our analysis, we assume a \textit{Euclid}-like source redshift distribution given by 
\citep[][]{blanchard2020euclid}{}{}:
\begin{align}
    n_\mathrm{s}(z) \propto \bigg(\frac{z}{z_0}\bigg)^2 \exp \bigg[-\bigg(\frac{z}{z_0}\bigg)^{3/2}\bigg],
\end{align}
with $z_0 = 0.9/\sqrt{2}$. 

\begin{figure}
	\includegraphics[width=\columnwidth]{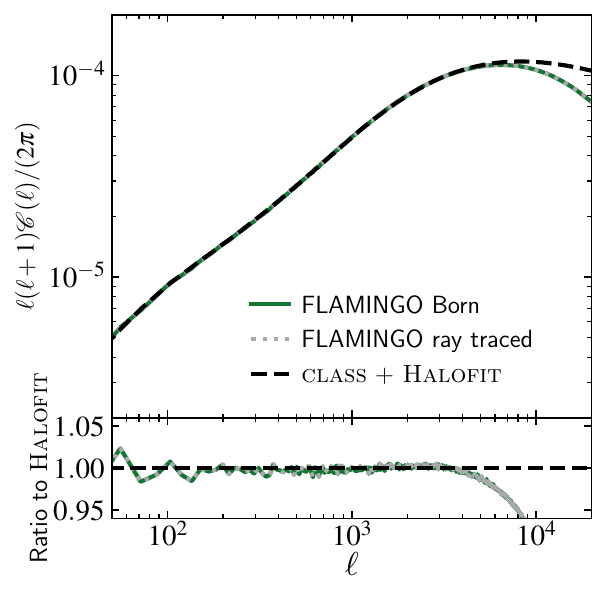}
    \caption{Top: Weak lensing convergence angular power spectrum estimated using a \textit{Euclid}-like source redshift distribution, corresponding to the mean of the 8 L2p8$\_$m9$\_$DMO observers. The full-sky map is constructed from the FLAMINGO mass maps and assumes the Born approximation (solid green) or a more advanced backward ray-tracing method (dotted gray). The dashed black curve shows the prediction from \textsc{Halofit} using the non-linear 3D matter power spectrum provided by \textsc{class}. Bottom: ratio to the \textsc{class} + \textsc{Halofit} prediction. The three estimates agree to an excellent degree apart from small scales due to the pixelisation of the HEALpix maps.}
    \label{fig:pow_spec_compare}
\end{figure}

We randomly rotate the mass maps whenever the light-cone diameter exceeds the box size to avoid encountering the same structure multiple times along the line of sight but, at the same time, not unnecessarily erasing correlations along the line of sight. In Appendix~A of \citet{Broxterman2024} we found that the rotation prescription had little impact on the final number density of WL peaks. To validate our map construction procedure, we first compare the angular power spectrum of the final WL convergence map to the prediction from \textsc{Halofit} \citep[][]{Takahashi2020Halofit}{}{} provided by \textsc{class} \citep[][]{blas2011cosmic}{}{} in Fig.~\ref{fig:pow_spec_compare}. \textsc{class} provides the three-dimensional matter power spectrum ($P_\mathrm{m}$), which we relate to the angular power spectrum ($\mathcal{C}(\ell)$) using \citep[][]{Limber1953approx,Loverde2008limber}{}{}:
\begin{align}
    \mathcal{C}(\ell) =  \frac{9 H_0^4 \Omega^2_{\mathrm{m}}}{4c^4}\int_0^{\chi_{\mathrm{hor}}} \mathrm{d} \chi\, \frac{W^2(\chi)}{\chi^2 a^2(\chi)} \,P_{\mathrm{m}}\bigg(\frac{\ell+1/2}{\chi},\, z(\chi)\bigg),
\end{align}
where $W(\chi)$ is given by equation~\ref{eqn:ana_lensing_kernel}. The \textsc{Halofit} + \textsc{class} prediction is given by the black dashed curve. The green solid line shows the estimate using the employed methods in this paper, i.e. assuming the Born approximation and using the \textsc{healpy} \textsc{anafast} routine. Finally, the dotted gray curve shows the angular power spectrum generated using the more elaborate backward ray-tracing method used in \citet{Broxterman2024}. The bottom panel shows the ratio to the \textsc{class} + \textsc{Halofit} prediction. At the power spectrum level, the ray tracing and Born approximation give the same results, as expected \citep[][]{hilbert2020accuracy}{}{}. The estimates also agree well with the \textsc{Halofit} prediction and illustrate that the map construction algorithm works well. The difference at small scales is caused by the pixelization of the matter field onto the HEALpix grid, as discussed in more detail in Section~3.5 of \citet{Broxterman2024}. For a detailed comparison between predictions from \textsc{Halofit} and FLAMINGO, see \citet{Upadhye2023}.

\subsection{Weak-lensing peaks}\label{sec:peaks}
To mimic a \textit{Euclid}-like signal, we smooth the final maps with a Gaussian beam with a smoothing scale of 1 arcmin and we apply galaxy shape noise by drawing from a normal distribution with mean $\mu = 0$ and standard deviation $\sigma = \sigma_\epsilon / \sqrt{2 n_{\mathrm{gal}} A_{\mathrm{pix}}}$, where $\sigma_\epsilon = 0.26$ is the rms total intrinsic ellipticity of the source galaxies, $n_{\mathrm{gal}} = 30$~arcmin$^{-2}$ is the source galaxy number density and $A_{\mathrm{pix}}$ is the area of a HEALpix pixel \citep[][]{Kaiser1993noise,vWaerbeke2000,Martines2019Euclidprep,Euclid2023HOWLS,EuclidPaper1_2024}{}{}.\footnote{This noise configuration likely slightly underestimates the actual noise that will be measured by Stage IV WL surveys but we still adopt it to facilitate a consistent comparison with \citet{Broxterman2024}.} By smoothing the maps with different smoothing lengths, information corresponding to different scales may be probed, as illustrated by \citet{li2019constraining}. In this work, we focus on a single smoothing scale but note that the main conclusions of the halo-peak matching are independent of the exact value of the smoothing scale, as illustrated in Appendix~\ref{app:smoothing}. We define a peak as any pixel on the HEALPix grid with a value greater than those of its 8 nearest neighbors. Although a crude definition, this has been used widely in observations \citep[e.g.][]{hamana2015rays,Shan2018,Grandon2024}{}{} as well as simulations \citep[e.g.][]{weiss2019peaks,coulton2020weak,Davies2024}{}{}.

To match the peaks to haloes, we assign two angular coordinates and a redshift to each peak and halo (see also Section~\ref{sec:halolc}). The positions of the peaks on the HEALPix grid directly provide the angular coordinates of the peaks. To assign a redshift to each peak, we exploit the summation properties of the Born approximation. As explained in Section~\ref{sec:mass_maps}, the WL signal is estimated with a double summation (equations~\ref{eqn:kappa_born_discreet}~\&~\ref{eqn:lensing_kernel_discreet}). The first summation sums over all the lens planes, and for each lens plane, the second summation calculates the contribution of all more distant source planes. For each peak, we determine the redshift of the shell where the contribution from the lens (equation~\ref{eqn:kappa_born_discreet}) is the largest and assign this redshift to the peak. Before determining the maxima, we smooth the contributions at each shell with the same smoothing kernel as applied to the final map. We refer to the redshift assigned in this way as $z_{\mathrm{true}}$ as this is the redshift of the major overdensity that acts as the lens to generate the peak. After matching the peaks to individual haloes, as explained in Section~\ref{sec:halo_peak_matching}, the peaks are assigned the redshift of the halo to which they have been matched. In Section~\ref{sec:obs_zs}, we will modify these redshifts to account for realistic observational redshift uncertainties.

The straightforwardness with which we can directly assign a redshift to each peak is the main reason why we choose to adopt the Born approximation. Assigning a redshift to a peak is less straightforward using the ray-tracing methodology used in \citet{Broxterman2024}, as that approach effectively sums over the contributions of all the sources, instead of the lenses.

\subsubsection{Purity}
To ensure that we select a peak sample that predominantly corresponds to a single halo along the line of sight towards each peak, we apply cuts based on the peak value, applied noise, and most and second most contributing lens shells, all measured at the angular position of the peak. First, if the smoothed random noise alone can account for over 50 per cent of the final peak value, we call this peak a `Noise peak'. Second, to guarantee that the object corresponds to a single shell and it is not a result of a superposition of multiple objects along the line of sight, we require the most contributing shell to contribute more than 50 per cent of the final peak value and the second most contributing shell to contribute less than 50 per cent of the most contributing shell. We refer to the subset of peaks satisfying these criteria as `Single shell peaks'. These cut will throw away a small subset of haloes located directly on the edge of two subsequent shells, which are, in reality, single haloes. As we aim to get a pure selection of peaks, we do not study these peaks in more detail, but we note that, as a consequence, our resulting selection of Single shell peaks will likely be slightly underestimated. We refer to the remaining set of peaks as `Multiple shells peaks', as their WL convergence cannot be attributed to the applied noise or a single shell, or there are multiple contributing shells along the line of sight.

\begin{figure}
	\includegraphics[width=\columnwidth]{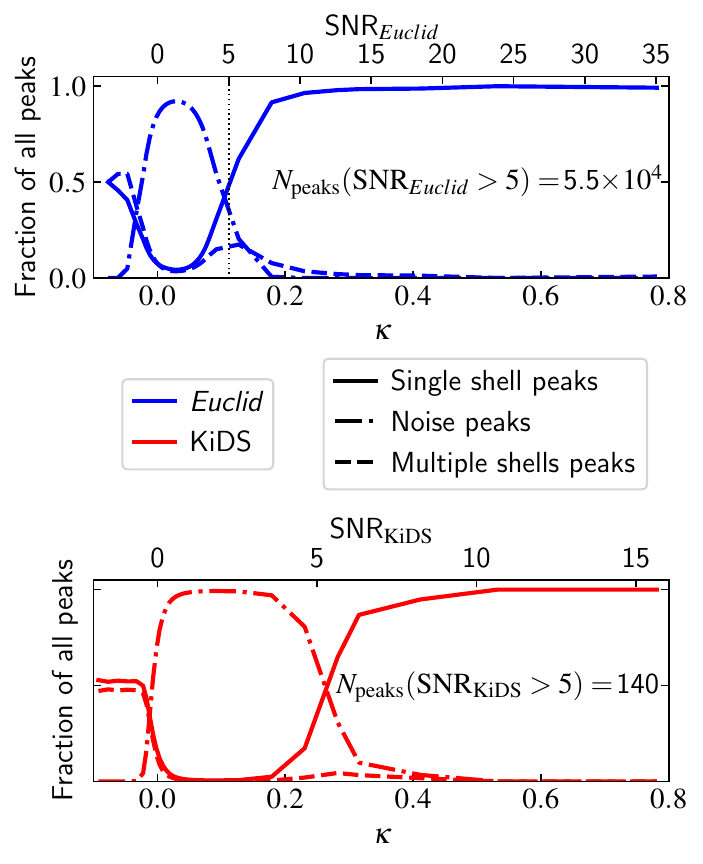}
    \caption{Fractional contribution of WL peaks that stem from a single redshift shell (`Single shell', solid), were assigned a random noise that can account for over half of the final peak signal (`Noise peaks', dashed-dotted), or originate from multiple shells along the line of sight (`Multiple shells peaks', dashed) as a function of WL convergence ($\kappa$). The blue and red curves correspond to a \textit{Euclid}- and KiDS-like analysis, respectively. The purity (fraction of Single shell peaks) increases with $\kappa$. We apply an SNR-cut of 5, as indicated by the vertical dotted line in the top panel, as selection criteria for the remaining analysis. The number of SNR~>~5 peaks estimated for the final data releases of the surveys is indicated in each panel, assuming an area of $1,500$~deg$^{2}$ and $14,000$~deg$^{2}$, for KiDS and \textit{Euclid}, respectively. Stage IV WL surveys will resolve a large amount of high-SNR peaks corresponding to massive clusters that could be used to constrain cosmology.}
    \label{fig:purity}
\end{figure}

In Fig.~\ref{fig:purity}, we show the fractional contribution of each of the three peak categories to the total number of peaks as a function of their WL peak convergence value. The top panel corresponds to a \textit{Euclid}-like analysis, whereas the bottom panel corresponds to a Stage III WL survey-like analysis. The top axis of each panel shows the SNR ($=\kappa/\sigma_{\mathrm{noise}}$). The dashed-dotted, dashed, and solid curves indicate the Noise, Multiple shells, and Single shell peaks, respectively. The curves correspond to the mean estimates of the eight observers in L2p8$\_$m9, but we find the same conclusions apply for the measurement in the DMO runs, i.e. baryonic effects do not change this picture.

Focusing first on the top panel, the highest-valued peaks nearly always belong to the Single shell category, as they are caused by a single overdensity along the line of sight. At $\kappa$ < 0.4 (0.2), the Multiple shells (Noise) peaks become increasingly important. Below SNR~=~5 ($\kappa<0.11$), the Noise peaks dominate the signal. This is expected as the noise value necessary to account for over half of the signal is increasingly less likely to be drawn for higher-valued peaks. For $\kappa \lesssim 0$ the Noise peak contribution drops again. In Appendix~\ref{app:noise_peaks}, we show in more detail that these peaks do not arise from the noise. We find that separating these negative-valued peaks into Single shell and Multiple shells peaks depends strongly on the exact choice of division criterion, which in this case is the ratio of the second to the most contributing shell and the ratio of the most contributing shell to the final peak value. However, this is only the case for the low-$\kappa$ peaks, and the $\kappa \gtrsim$ 0.1 subset, which contains the WL peaks considered in the remainder of this work, is robustly split into the different categories. The contribution of Multiple shells peaks is $\sim10$ per cent at SNR~=~5 and decreases monotonically until it vanishes at SNR~=~18 ($\kappa=0.40$).

These results agree with those from \cite{Hamana2004}~\&~\citet{yang2011cosmological} who used a different approach but also found that high-signal peaks (in their case SNR~>~3.5) principally stem from a single massive halo each. In Fig.~\ref{fig:purity}, we see that their convergence cuts are in the regime where the purity drops steeply as the noise and superposition of multiple objects along the line of sight generate a larger fraction of the peaks for lower $\kappa$ values. At this point, we have not yet aimed to match individual peaks to haloes, but based on the results of \cite{Hamana2004,yang2011cosmological,liu2016origin}, we expect that we should be able to match most of the peaks to a single massive halo each, as we will indeed show in Section~\ref{sec:halo_peak_matching}. 

To illustrate the major improvement of the Stage IV WL surveys, we compare the \textit{Euclid} predictions to those resulting from using the same methods but assuming Stage III characteristics. The red curves in the bottom panel of Fig.~\ref{fig:purity} illustrate the same 3 peak categories, but are based on a WL map constructed using the KiDS DR4 (KiDS-1000) source redshift distribution and noise properties \citep[][]{Kuijken2019KiDS,asgari2021kids}{}{}. The KiDS signal is dominated by Noise peaks to larger $\kappa$ values, and therefore, fewer peaks can be confidently assigned to a single halo. The main difference rests in the source galaxy background number density, which is a factor $\approx 5$ greater for \textit{Euclid}, and therefore, the mean noise is a factor $\approx 2.2$ times greater for the KiDS analysis. Additionally, Stage IV surveys detect more distant galaxies, so the WL signals will also be stronger, while the area of the \textit{Euclid} survey will be $\sim$10 times greater, which impacts the total number of measured objects. 

Selecting a higher WL convergence cut gives a purer sample as the fraction of Single shell peaks increases. For an SNR-cut of 5 (8), we find a purity percentage (percentage of Single shell peaks) of 76 (96) per cent for \textit{Euclid} and 56 (100) per cent for KiDS, assuming a KiDS-like noise such that SNR~=~5 corresponds to $\kappa~=~0.25$ for the latter predictions. We stress that these percentages are likely underestimated due to the conservative cuts we have applied to assign a peak to a category. Not only the purity of the \textit{Euclid} sample is higher at a fixed WL convergence cut, the total number of peaks with SNR~>~5 (8) increases from 140 (4) in the full KiDS footprint ($\approx 1,500$~deg$^{2}$) to $5.5\times10^{4}$ ($6.8\times10^{3}$) in the full \textit{Euclid} footprint ($\approx 14,000$~deg$^{2}$), illustrating the major improvement of Stage IV surveys as they will measure over two orders of magnitude more high-valued peaks. The KiDS estimate is roughly consistent with the recent HSC Y3 analysis results from \citet{Chen2024} and \citet{Chiu2024}, who find 129 SNR~>~4.7 peaks on the WL aperture mass maps. While their area is smaller than the area we assume for the KiDS-1000 analysis, their survey is deeper, so we expect that their number roughly agrees with our simple KiDS estimate. 

We compare our purity prediction to those from \citet{yang2011cosmological}. They report a purity of only 20 per cent for SNR~>~3.5 peaks. This is 4 times lower than our purity, but lowering the SNR threshold greatly impacts the purity. Assuming their SNR cut, we still find a purity of 38 per cent. Their method attributes a peak to a halo by assuming NFW profiles for each simulated halo and calculating the expected convergence signal using this profile. They then match peaks and haloes using a maximum angular separation of 1.8 arcmin. If the estimated WL convergence is more than half the final WL peak value, they attribute the peak to a single massive halo. However, they already note that this approach's estimated WL convergence has a fractional bias of 7 per cent and underestimates the WL convergence values. As we do not make any assumptions about the halo profile and the division criteria differ, some differences between the two approaches are expected. In conclusion, we find similar purity percentages to those in previous work.

\citet{Hamana2004} estimate that the mean number of haloes that cause a WL peak with SNR~>~4 is 37 per deg$^2$. At the same SNR cut, we find a lower peak number density of only 11 peaks per square degree. Based on the source redshift distribution, we expect their signal to be stronger than we measure. With their SNR cut, they report a contamination rate, i.e. a percentage of peaks that cannot be attributed to a single halo, of 44 per cent. The main difference between their analysis and ours is the choice of SNR cut. From Fig.~\ref{fig:purity} it is clear that the noise and superposition contributions vanish quickly around this SNR regime. If we apply their SNR cut, we find a similar contamination rate of 47 per cent.

In the remainder of our analysis, we apply an SNR cut to our peak sample, as indicated by the dotted black curve in the top panel of 
Fig.~\ref{fig:purity}. We will only use peaks with an SNR value greater than the cutoff value of SNR~=~5 to determine the redshift evolution and perform the halo-peak matching. We choose this value because at larger SNR, Single shell peaks dominate and we thus have a reasonably pure sample. Additionally, we stress that until now, we have been conservative in our noise and multiple shells cuts as we find that most SNR~>~5 peaks belonging to either of these categories still clearly correspond to a single object in the overdensity field, as illustrated in more detail in Appendix~\ref{app:extra_field}. When studying the redshift distributions of the peaks, we assign a redshift to all Noise and Multiple shells peaks above the SNR-cut and include them in the analysis, as in an observational campaign there is no way of deciding which peaks are generated by the noise.

In \citet{Broxterman2024} we argued that, for a similar analysis, using the number density of WL peaks as a function of their WL convergence strength, the $\kappa$ = [0.1,0.4] regime is most useful for WL peak inferences, as the differences are most easily interpretable in that range. Lower-valued peaks require accurate modeling of instrument and survey characteristics, while the number of higher-valued peaks is subject to significant cosmic variance. Additionally, for this intermediate regime, the differences between the cosmology and baryonic variations could be easily understood by considering the impact of baryonic physics on single haloes and the impact of different cosmological parameters on the HMF. The peaks belonging to this WL convergence regime are predominantly included in our sample after the SNR~>~5 cut.

\subsection{Halo light-cone}\label{sec:halolc}
To match the peaks to haloes, we use the FLAMINGO halo light-cones. FLAMINGO uses the halo finder HBT-HERONS to identify haloes and track their formation and evolution, including mergers \citep[]{Han2018,moreno2025assessingsubhalofinderscosmological}{}{}. The halo finder is used to identify central haloes and their most-bound BH particle, which is set as the centre of the halo. The 3-dimensional positions of the haloes are then converted to angular positions and redshifts, as seen by the different virtual observers in the runs, and saved as halo light-cone files. We directly use these files for our halo-peak matching. We rotate the angular coordinates of the haloes using the same rotations as the light-cone mass maps. On top of the halo finder, which determines the centres of the haloes in the simulation, the FLAMINGO data includes a large set of precomputed quantities in Spherical Overdensity APerture (SOAP) catalogues (McGibbon et al. in prep). From the SOAP catalogues, for each halo, we use $M_{\mathrm{200c}}$, the mass within a spherical aperture with radius $R_\mathrm{200c}$ enclosing an average overdensity of $\langle \rho (z) \rangle = 200 \rho_\mathrm{crit}(z)$, of the closest snapshot in redshift. 

\subsection{Halo - peak matching}\label{sec:halo_peak_matching}
Using the angular coordinates and redshifts of the peaks and haloes, we now match individual peaks to haloes from the FLAMINGO halo light-cones. We perform the matching separately for each shell.\footnote{We find that allowing a peak to be matched to a halo in the previous or subsequent shells if the haloes are close to the edge of the shell improves a small number of matches. In these cases,  on one side of the edge, there is a high-valued WL peak without a corresponding massive halo. On the other side, at the same angular position, there is a massive halo without a corresponding peak. We allow a peak to be matched to haloes that are within 1~cMpc of the shell edge, which corresponds to $R_\mathrm{200c}$ of a $M_{\mathrm{200c}} \approx 10^{14}~\mathrm{M_\odot}$ halo. Doubling this value does not significantly change the results.} For each shell, we select all peaks corresponding to that shell. As these peaks all have the same redshift (i.e. the central redshift of the shell), their WL kernel (equation \ref{eqn:lensing_kernel_discreet}) is unchanged, and a higher peak should directly originate from a more overdense structure. However, as there may be small variations due to other minor over- or underdense regions within the shell or along the line of sight, the noise that is applied in the end, and because the haloes are not perfectly spherical NFW haloes but differ in shape \citep[e.g.][]{Velliscig2015,Chua2019}{}{}, these peaks will not all correspond to the most massive haloes. Therefore, for each shell, we select the $N$ most massive haloes such that $N_{\mathrm{halo}} = 100 N_{\mathrm{peak}}$, which are then allowed to be matched to the peaks. We find that changing this value by a factor of two does not impact the results.

To match the haloes to the peaks, we determine the great circle distance ($\Delta \theta$) between each peak and all nearby haloes. First, we project the haloes onto the HEALPix pixels according to their location on the halo light-cone. Then, using the \textsc{healpy} function \textsc{query$\_$disc}, we select all haloes within 2 degrees of the peak. We then calculate the distance between the peak and all haloes in the angular aperture and scale the distances by the quantity $\theta_\mathrm{200c} = R_\mathrm{200c}/d_\mathrm{A}$, which is a proxy for the angular size of the halo on the sky, where $d_{\mathrm{A}}$ is the angular diameter distance. Then, we select the closest halo in $\Delta \theta/\theta_\mathrm{200c}$ for each peak.\footnote{We find the results do not change if we instead match peaks to haloes based on $\Delta\theta / L_{*,50\mathrm{kpc}}$ or $\Delta\theta / M_{*,50\mathrm{kpc}}$, where $L_{*,50\mathrm{kpc}}$ and $M_{*,50\mathrm{kpc}}$ are the stellar luminosity in the GAMA $r$ band and stellar mass in a 50 kpc aperture, respectively.} We do not require a bijective match between the halo and peak as we find that a single halo may cause multiple peaks in the final map, particularly for massive, low-redshift, or merging systems. If, instead, we match peaks to the closest halo in $\Delta \theta/\theta_\mathrm{200c}$ without first assigning the peak to a redshift, i.e. matching to all haloes in the same angular region we find that 95 (97) per cent of the SNR~$>$~5 (8) peaks are matched to the same haloes. The comparison indicates minor dependence on the matching algorithm due to projection effects but we find this does not impact our main conclusions.

\begin{figure*}
	\includegraphics[width=\textwidth]{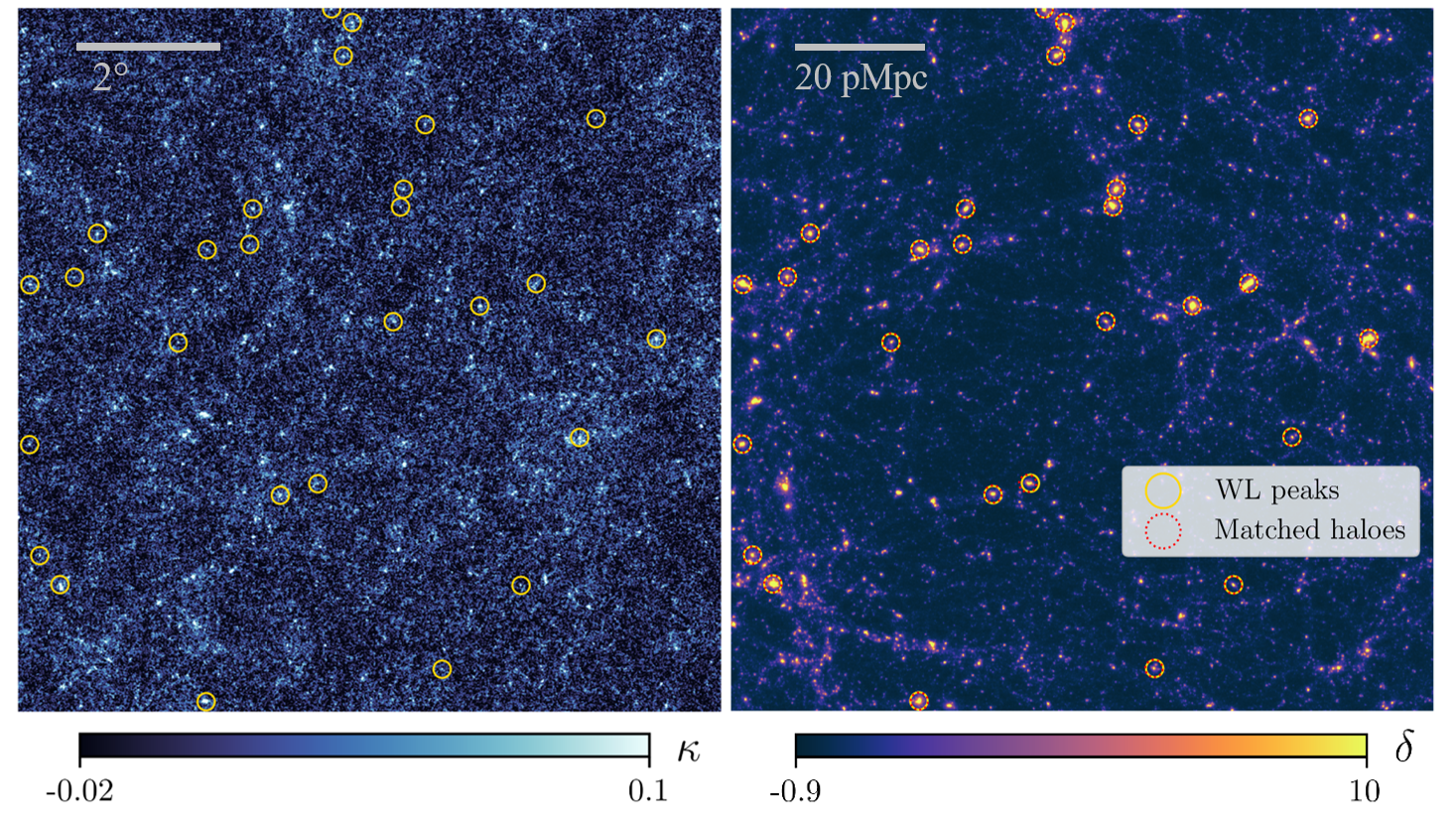}
    \caption{Left: 10$\times$ 10 deg$^2$ patch of the \textit{Euclid}-like WL convergence field of the virtual observer in L1$\_$m9. The WL peaks are selected from this map. The SNR~>~5 peaks that are assigned to the fourth shell ($z = [0.15-0.2]$) are indicated by the yellow circles. Right: projected overdensity field of the fourth shell of the same angular field, corresponding to a physical size of $\approx 110 \times 110$~pMpc$^2$. The yellow circles correspond to large overdensities. Red dotted circles show the positions of the haloes to which the peaks are matched. Most of the visibly overdense structures with no associated peak generate a peak with a lower SNR-value than the cut we apply, as illustrated in Appendix~\ref{app:extra_field}.}
    \label{fig:gnomview_plot}
\end{figure*}

Fig.~\ref{fig:gnomview_plot} illustrates the accuracy of the matching procedure. The left-hand panel shows a 10 $\times$ 10 square degree patch from the final WL convergence map of the virtual observer in L1$\_$m9 from which the peaks are selected. In this panel, the SNR~>~5 peaks that correspond to the fourth shell ($z = [0.15-0.2]$) are indicated by yellow circles. The bright peaks that do not have a circle have been assigned to other redshift shells. The right-hand panel shows the overdensity field of the fourth shell for the same angular region. The field corresponds to a physical size of $\approx 110 \times 110$~pMpc$^2$.  Each yellow circle assigned to this shell visually corresponds to an overdense structure. The red dotted circles indicate the positions of the haloes to which the peaks have been matched. All peaks and haloes are at almost the same position, indicating the peaks are matched correctly. 

Some peaks, such as the one to the left of the legend, have a small angular offset compared to the halo centre. As no other haloes are near the same position, we are still confident that the peaks have been matched to the correct haloes. Minor differences between the halo light-cone coordinates and smoothed projected WL overdensity are expected, as we apply noise and smooth the final maps and because of projection effects. Other structures seemingly equally overdense as the ones that are associated with a WL peak do not have a corresponding peak. Most of these structures cause a peak in the final WL convergence map with an SNR value below the cut-off value we apply, as illustrated in Appendix~\ref{app:extra_field}. Alternatively, a peak at that position has been assigned to a more strongly contributing structure along the line of sight. As we chose to keep the peak catalogue relatively pure, we do not aim to include these peaks in this analysis. Still, we stress that the SNR cut we applied has not been optimised and may be lowered to include more peaks, albeit at the cost of purity. 

In Appendix~\ref{app:extra_field}, we split the peaks in Fig.~\ref{fig:gnomview_plot} into the three categories considered in this work (i.e., Single shell, Noise, and Multiple shells peaks). There, we show that, generally, Noise and Multiple shells peaks are still clearly assigned to a massive halo. In that appendix, we also show the 3~<~SNR~<~5 peaks that were assigned to this shell and illustrate that most overdense structures that do not have a corresponding peak in Fig.~\ref{fig:gnomview_plot} generate a peak with an SNR-value below the cut that we apply.

\subsection{Observed redshifts}\label{sec:obs_zs}
Until now, we have used the simulation properties to identify structures with high precision and accuracy at any point in space and time within the simulation. However, as observations typically assign a redshift based on a limited number of photometric bands, we explore how the expected associated observational uncertainties impact our results.

\textit{Euclid} will rely on photometric redshifts for the bulk of the measured galaxies. The survey requirement is to measure the photometric redshifts such that the standard deviation on the uncertainty satisfies $\sigma_z < 0.05(1+z)$ with a catastrophic failure rate of less than 10 per cent \citep[][]{EuclidPaper1_2024}{}{}. However, the peaks we measure correspond to the most massive clusters, as we will show in Section~\ref{sec:peak_in_M200}, which will likely contain bright and red objects that will be easier to identify using either the regular methodology or a red-sequence cluster identification mechanism such as redMaPPer \citep[][]{Rykoff2014}{}{}. Therefore, assuming the general \textit{Euclid} requirement will overestimate the redshift uncertainty on our peak sample. The assumption that the WL peaks can be associated with optical clusters using red-sequence-based cluster finders is confirmed by the recent HSC Y3 analysis of \citet{Chen2024} and \citet{Chiu2024}, who find that they can match all their SNR~>~5 peaks selected from aperture mass maps to an existing measurement of an optical cluster from CAMIRA \citep[][]{Oguri2018}, redMaPPer or WHT \citep[][]{Wen2015}.

The \textit{Euclid} satellite will measure photometry in 1 visual ($m_\mathrm{VIS}$) and 3 near-infrared ($YJH$) bands, but the photo-$z$ measurements will strongly rely on ground-based photometry. The satellite's measurements will be complemented by 6 ground-based photometric measurements in the $ugrizy$ bands from a range of different facilities \citep[][]{EuclipPrep12,EuclidPaper1_2024}{}{}. As a more accurate estimate of the redshift uncertainty, we take as a reference the KiDS bright sample, which is a bright subsample of $\sim 10^6$ galaxies of KiDS-1000 that uses the 9-band $ugriZYJHK_s$ photometry to determine photo-$z$s \citep[][]{Kuijken2019KiDS,Bilicki2021}{}{}. By comparing to the GAMA spectroscopic redshifts, they find they assign photo-$z$s to the bright subsample with an accuracy of $\sigma_z = 0.018(1 + z)$ up to $z=0.5$ without catastrophic failures. \citet{Vakili2023} select a KiDS-1000 subsample using a red sequence template that includes estimating the red sequence redshifts, and by comparing to spectroscopic redshifts, they find similar accuracy and precision as the KiDS bright sample extending up to $z=0.8$.

As the number of bands and depth used for the \textit{Euclid} photometric redshift estimates will be similar to KiDS, we expect that the photometric redshifts for the clusters that correspond to our WL peaks will be measured with similar accuracy and precision as the KiDS bright sample. To mimic the observed redshifts, for each $z_\mathrm{true}$ that has been assigned to a peak, we therefore draw an observed redshift according to a normal distribution with mean $\mu$ and standard deviation $\sigma$,
\begin{align}\label{eqn:ztrue2zobs}
    \mathcal{N} \bigg\{ \mu = z_{\mathrm{true}} ,  \sigma_z = 0.02(1+z_{\mathrm{true}}) \bigg\},
\end{align}
which is truncated at $z=0$. We will only show the redshift distributions of all the baryonic and cosmology variations as a function of the mock photometric redshifts $z_{\mathrm{obs}}$. Still, we note that the same conclusions hold if we instead use the true redshifts or modify the redshifts with an uncertainty of $ \sigma_z = 0.05(1+z_{\mathrm{true}})$ and an extreme outlier fraction of 10 per cent, to the degree of the \textit{Euclid} requirements.

\section{Results}\label{sec:results}
We now present the results of the halo peak matching by showing the WL peak distribution binned by halo mass in Section~\ref{sec:peak_in_M200}. We then quantify the completeness of the matched haloes in Section~\ref{sec:completeness} and study the redshift distribution of the WL peaks in Section~\ref{sec:kappa_binned_in_z}. We quantify the degree of numerical convergence and cosmic variance in Sections~\ref{sec:numerical_convergence} and \ref{sec:cosmic_variance}, respectively. Finally, we qualitatively and quantitatively compare the difference between the cosmological and baryonic impact on the peak statistics in Sections~\ref{sec:z_evolution} and \ref{sec:ks_test_sec}. 

\subsection{Peak distribution in halo mass}\label{sec:peak_in_M200}
\begin{figure}
	\includegraphics[width=\columnwidth]{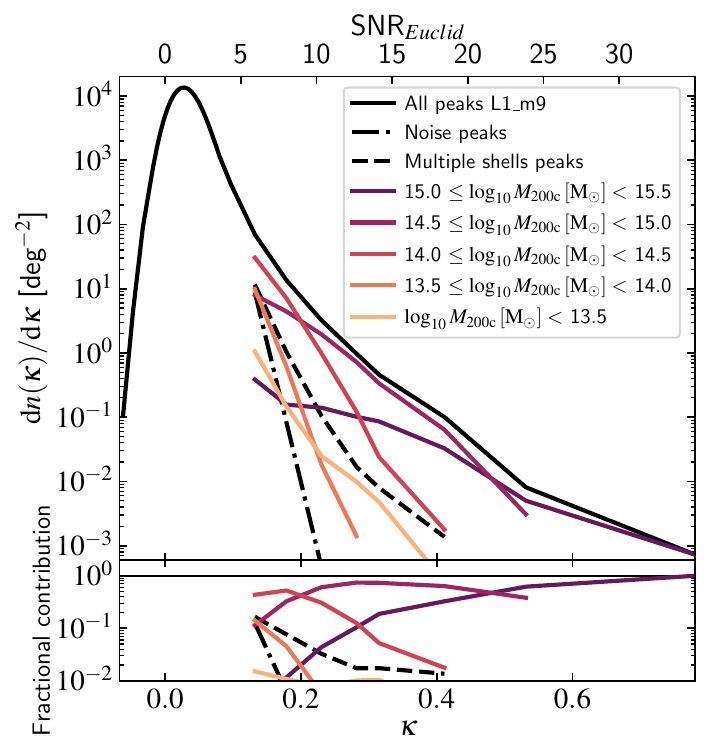}
    \caption{Top: number density of WL peaks as a function of WL convergence (bottom x-axis) and SNR (top x-axis) for the virtual observer in L1$\_$m9 (solid black). Noise and Multiple shells peaks are indicated by black dashed-dotted and dashed curves, respectively. Peaks with SNR~>~5 are binned by $M_{\mathrm{200c}}$ of the haloes to which they have been matched, with larger mass bins indicated by increasingly darker solid curves. Bottom: fractional contribution of all mass bins to the total number of haloes. The high-valued ($\kappa>0.11$; SNR~>~5) WL peaks primarily trace haloes with $M_\mathrm{200c} > 10^{14}~\mathrm{M_\odot}$.}
    \label{fig:halo_mass_binning}
\end{figure}

First, we show the results of matching the haloes to peaks by binning the peaks by halo mass. Fig.~\ref{fig:halo_mass_binning} shows the number density of WL peaks as a function of WL convergence for all redshifts combined. The top x-axis shows the SNR for the \textit{Euclid}-like analysis. The solid black curve gives the distribution of all peaks measured in L1$\_$m9. The SNR~>~5 selection, which primarily corresponds to a single halo per peak, is binned by the halo masses of the haloes to which the peaks have been matched, as depicted using increasingly darker colors for the larger halo mass bins. The figure also shows the Noise peaks (dot-dashed) and Multiple shells peaks (dashed), as detailed in Section~\ref{sec:peaks}. The fraction of Noise peaks decreases with increasing WL convergence. At SNR~=~5, the Noise peaks account for $\sim 10$ per cent of all peaks, but this quickly decreases to 0 at SNR~=~10 (see also Fig.~\ref{fig:purity}). 

The bottom panel of Fig.~\ref{fig:halo_mass_binning} shows the fractional contribution of all mass bins. As expected, we find that, on average, higher-valued WL peaks correspond to more massive haloes. We find qualitatively similar results to \citet{Hamana2004, yang2011cosmological, liu2016origin}, who found SNR~>~3.5 peaks are predominantly caused by a single massive halo each. We can now study the peaks as a function of their halo mass in more detail. For $\kappa > 0.25$, we find that over 95 per cent of the peaks can be attributed to haloes with $M_{\mathrm{200c}} > 10^{14}$~$\mathrm{M_\odot}$. Of all SNR~>~5 peaks, the majority are matched to haloes with $M_\mathrm{200c}~=~10^{14} - 10^{14.5}$~$\mathrm{M_\odot}$. In \citet{Broxterman2024}, we already hypothesised that the WL peaks originate primarily from $M_{\mathrm{200c}} > 10^{14}$~$\mathrm{M_\odot}$ haloes based on the differences between the thermal and kinetic jet AGN variations and the comparison of the HMFs of these runs. We now confirm this hypothesis. 

\citet{liu2016origin} used observational data from the CFHTLens survey to link haloes to WL peaks. They found that each peak with SNR~>~3.5 corresponds to a single halo of $M_{\mathrm{vir}}\sim10^{15}~\mathrm{M_\odot}$. Qualitatively, this agrees with our findings as their analysis probes less deep, but our highest-valued $\kappa$ peaks are dominated by haloes with $M_\mathrm{200c} > 10^{15}~\mathrm{M_\odot}$.

\subsection{Completeness}\label{sec:completeness}

\begin{figure}
	\includegraphics[width=\columnwidth]{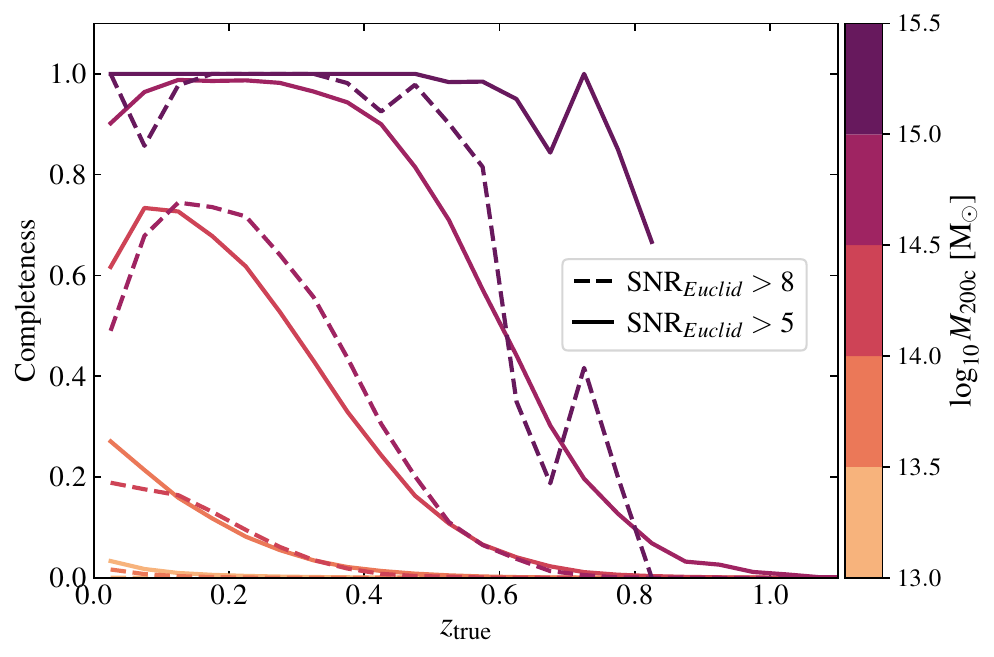}
    \caption{The completeness, i.e. the fraction of haloes matched to a high-valued WL convergence peak, as a function of the halo redshift. The haloes are binned by $M_{\mathrm{200c}}$, and more massive halo mass bins are shown in increasingly darker colors. The solid curves correspond to the fiducial SNR~>~5 sample, while the dashed curve only selects haloes matched to peaks with SNR~>~8.  The completeness drops for lower masses and higher redshifts. The impact of the smoothing scale on the completeness is studied in more detail in Appendix~\ref{app:smoothing}.}
    \label{fig:completeness_with_z}
\end{figure}

We define completeness as the fraction of haloes within a halo mass bin matched to at least one WL peak over the total number of haloes in that bin. Fig.~\ref{fig:completeness_with_z} shows the completeness as a function of redshift and binned by halo mass. The solid curves correspond to haloes matched to peaks with SNR~>~5, and the dashed curves include only those matched to SNR~>~8 peaks. The cumulative completeness up to $z = 0.5$ and $1.0$ for haloes matched to SNR~>~5 peaks is given in Table~\ref{tab:completness}. The table also includes the total number of haloes per square degree on the halo light-cone until $z=3$. We find excellent completeness for the most massive haloes. Of all 929 $M_{\mathrm{200c}} > 10^{15.0}~\mathrm{M_\odot}$ haloes, 98 per cent has been matched to an SNR~>~5 peak. Up to $z = 0.5$, this fraction is 100 per cent.  Of all $5.0 \times 10^4$  $M_{\mathrm{200c}} = 10^{14.5} - 10^{15.0}~\mathrm{M_\odot}$ haloes, 52 per cent has a corresponding SNR~>~5 peak. Until $z=0.5$, this is 93 per cent. At higher redshifts, as well as for lower masses, the completeness drops. This is expected, as lower mass objects, on average, generate weaker WL signals, higher-redshift objects have fewer source galaxies to lens, and the contribution of the WL kernel decreases at larger redshifts. 

For the $10^{14} < M_{\mathrm{200c}} [\mathrm{M_\odot}] < 10^{15}$ bins, we predict decreasing completeness at low redshift ($z < 0.2$), which can be attributed to the fixed smoothing scale that is applied. For some of these objects, the smoothing scale is too small to achieve the optimal SNR, as they subtend a larger angular scale than is smoothed over. We illustrate this in more detail in Appendix~\ref{app:smoothing}, where we show the completeness increases for these low-redshift bins when adopting a larger smoothing scale. The most massive haloes are so massive that they are matched to a WL peak independent of the smoothing scale we use. As the redshifts of the peaks are not known before detecting them, a smoothing scale has to be chosen without considering that information. The adopted smoothing scale is a trade-off between higher completeness at lower or higher redshifts for a larger or smaller smoothing scale than the fiducial value adopted in this work, respectively. We adopt a 1 arcmin scale to facilitate the comparison with \citet{Broxterman2024}, but multiple scales may be combined to extract more information \citep[][]{liu2015cosmology}{}{}. 

The dashed curves in Fig.~\ref{fig:completeness_with_z} illustrate that the completeness drops when choosing a higher SNR threshold for the WL peaks. The increase in SNR-cut impacts the completeness of the most massive bin the least, as these haloes generally cause the highest-valued WL peaks.

\begin{table}
\centering
\caption{The completeness, i.e. the fraction of haloes matched to an SNR~>~5 WL peak, assuming a smoothing scale of 1 arcmin. This peak sample has a purity of 76 per cent (see Fig.~\ref{fig:purity}). The columns list the halo mass bin, $M_{\mathrm{200c}}$; the total number of haloes per deg$^2$ on the entire halo light-cone until $z=3$; completeness up to $z = 0.5$ and $1.0$.}
\label{tab:completness}
\begin{tabular}{llll}
\hline
$\log_{\mathrm{10}}M_{\mathrm{200c}} [\mathrm{M_\odot}]$ & No. haloes& Completeness & Completeness\\
 & [deg$^{-2}$] &  $z <0.5$ &  $z < 1.0$\\
\hline
$13.0 - 13.5$      & $8.9 \times 10^2$ & 0.002 & 0.001\\
$13.5 - 14.0  $    & $1.6 \times 10^2$ & 0.038 & 0.010\\
$14.0 - 14.5   $   & $1.9 \times 10^1$ & 0.378 & 0.136\\
$14.5 - 15.0  $    & $1.2 \times 10^0$ & 0.930 & 0.578\\
$15.0 - 15.5   $   & $2.3 \times 10^{-2}$ & 1 & 0.984 \\
\hline
\end{tabular}
\end{table}

\citet{Hamana2004} report their completeness in terms of the signal expected for a halo of mass similar to that in their simulation, but with an NFW profile instead of the actual profile of the halo they study. They report a completeness of 81 per cent for haloes with SNR$_{\mathrm{NFW}}$~>~5. In our analysis, instead, we illustrate the completeness binned by halo mass. However, at the same time, they reported an efficiency, i.e. a purity, of 82 per cent using this SNR$_{\mathrm{NFW}}$ cut. We find similar, although slightly better, values based on our methodology. Combined with the purity shown in Fig.~\ref{fig:purity}, we find that an SNR~>~5 cut will result in a pure sample (76 per cent) while measuring a large number of peaks (5.5$\times$10$^{4}$). We repeat that extending this to lower SNR is likely possible, as we find that most Noise and Multiple shells peaks are still attributable to a massive object.  

\subsubsection{Comparison to X-ray and SZ cluster abundance inferences}

Next, we compare our completeness to the state-of-the-art X-ray survey eROSITA. \citet{Ghirardini2024Erosita} use the cosmology subsample of the X-ray selected eRASS1 cluster sample in their cluster abundance inference. Their sample, consisting of 5,263 clusters on $\sim 13,000$~deg$^2$, has an estimated purity of 94 per cent \citep[][]{Bulbul2024}{}{}. In our analysis, we find that an SNR~>~8 WL peak sample, corresponding to $\sim 6,800$ WL peaks for \textit{Euclid} DR3, has a purity of 96 per cent, illustrating that WL peaks are competitive with state-of-the-art X-ray cluster cosmology inferences. Lowering the SNR cut to 5 results in an even greater difference in the number of probed clusters, as $N_{\mathrm{peaks}}(\mathrm{SNR} > 5)=5.5\times10^{4}$, but this is at the cost of the purity which decreases to 76 per cent. Comparing our completeness to that of the eROSITA sample reported in Appendix~D of \citet{Ghirardini2024Erosita}, we see that at $z = 0.3$, their sample is $\approx100$ per cent complete for $M_{\mathrm{500c}} > 10^{15.2}~\mathrm{M_\odot}$. At the same redshift, as illustrated in Fig.~\ref{fig:completeness_with_z}, we find our sample (with an SNR~>~5 cut) similarly complete for $M_{\mathrm{200c}} > 10^{14.5}~\mathrm{M_\odot}$. For SNR~>~8, at a similar redshift, our sample has a similar completeness level for $M_{\mathrm{200c}} > 10^{15.0}~\mathrm{M_\odot}$. For $z >0.6$, the completeness of our WL peak sample drops steeply as a consequence of the WL kernel, even for the most massive haloes. 

We also compare our results to the SZ selected samples from \citet{Hilton2021}. Their sample consists of 4,195 optically confirmed clusters over $\sim 13,000$~deg$^2$ extending up to a redshift of 1.9. At $z=0.5$, they report a completeness of 90 per cent down to masses of $M_\mathrm{500c} > 3.8 \times 10^{14}~\mathrm{M_\odot}$, which decreases for lower and higher redshifts as well as for lower masses. They expect 34 per cent of their candidates to be false positives, corresponding to a purity of 0.66. For $z>0.6$, the completeness of their SZ-selected sample is generally better, whereas our WL sample has better completeness for lower redshifts. 

To conclude, the SNR~>~5 WL peak sample that Stage IV WL surveys will measure corresponds to massive clusters and has a purity and completeness equal to or better than state-of-the-art X-ray or SZ cluster abundance inferences. This especially holds for redshifts up to $z\approx 0.7$ and may be improved further by combining smoothing scales as illustrated in Appendix~\ref{app:smoothing}. At the same time, WL peaks circumvent some of the selection effects and model assumptions, as both SZ and X-ray analyses require assumptions on the parametric form and scatter in the scaling relation relating the observable to the halo mass. Compared to an X-ray or SZ cluster analysis, the limiting factor of this analysis is that the WL kernel restricts the accessible redshift range. Therefore, the redshift range that we probe does not extend to as high a redshift as is probed by X-ray or SZ cluster analyses. In Section~\ref{sec:peak_in_M200} we have shown that the SNR~>~5 peaks probed by stage IV WL surveys primarily trace $M_{\mathrm{200c}} > 10^{14}~\mathrm{M_\odot}$ haloes. Physically, these correspond to massive clusters of galaxies that should be visible in the optical, SZ, and X-ray \citep[e.g.][]{Allen2011}. The analysis of this paper could, therefore, possibly be extended by including SZ-selected clusters for higher redshifts \citep[e.g.][]{Bocquet2023SPTwlensing1,Bocquet2024SPTwlensing2}.

\subsection{Redshift distribution of SNR > 5 WL peaks}\label{sec:kappa_binned_in_z}
In this section, we study the redshift distribution of the SNR~>~5  WL peaks. We stress that all SNR~>~5 peaks, including the Noise and Multiple shells peaks, are assigned a redshift. First, we show the number density of WL peaks binned by redshift. We then investigate the degree of numerical convergence and cosmic variance. Finally, we study the redshift distributions as a function of mock photometric redshifts to quantify the impact of baryonic physics and cosmology using the different FLAMINGO 1~cGpc variations.

\begin{figure}
	\includegraphics[width=\columnwidth]{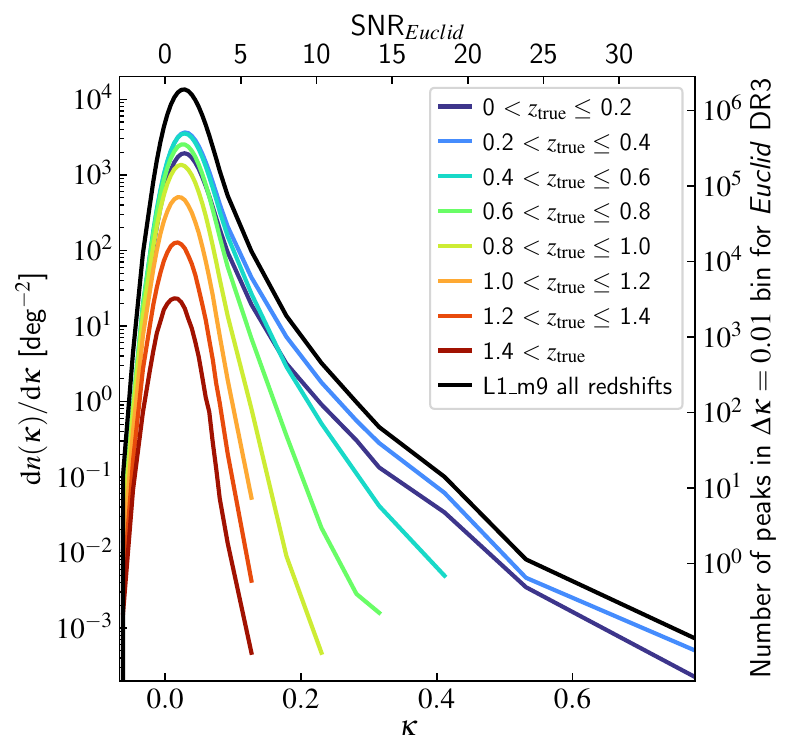}
    \caption{Number density of WL peaks as a function of WL convergence binned by redshift for L1$\_$m9. The redshift is assigned by identifying the lensing plane that contributes most to the final WL peak convergence value (Equation~\ref{eqn:kappa_born_discreet}). The redshift distribution is determined by the WL kernel, which is set by the source redshift distribution, and the evolution of structure formation. The distribution peaks at z = [0.2-0.4] as expected from the WL kernel peaking at $z \approx 0.375$. The right axis indicates the number of WL peaks in a $\Delta \kappa = 0.01$ bin for \textit{Euclid} DR3.}
    \label{fig:number_dens_binned_in_z}
\end{figure}

Fig.~\ref{fig:number_dens_binned_in_z} illustrates the number density of WL peaks as a function of WL convergence binned by their redshift. The right-hand axis shows the number of peaks that we estimate will be detected in a $\Delta \kappa = 0.01$ bin in the final \textit{Euclid} data release. As explained before, the redshift distribution of high-$\kappa$ WL peaks depends on the evolution of structure formation and on the WL kernel set by the source galaxy redshift distribution. However, it is also sensitive to cosmology due to the comoving volume-redshift relation. Therefore, the WL peak distributions contain both geometrical and perturbative information. 

Fig.~\ref{fig:number_dens_binned_in_z} illustrates that, generally, there are no peculiar features in the $\kappa$ distribution of the peaks with different redshifts other than overall amplitude differences. We find that peaks are more likely to be generated from lower redshift bins (particularly $z = [0-0.6]$), as this is where most of the structure has formed and the WL kernel peaks for the \textit{Euclid}-like source distribution, independently of the WL convergence value. Only the lowest redshift bin, $z = [0-0.2]$ (dark blue), shows different behaviour, as it contains fewer $\kappa \lessapprox 0.15$ peaks than in the $z=$ [$0.4-0.6$] and [$0.6-0.8$] bins, but for larger WL convergence values, it has more peaks. This characteristic originates from a subset of the high-mass haloes at low redshifts that generate multiple WL peaks. As discussed in the previous section, at low redshifts, the applied smoothing scale is smaller than the angular scale subtended by the most massive objects. These objects may therefore generate multiple peaks, which is reflected in Fig.~\ref{fig:number_dens_binned_in_z} as the lowest redshift bin, compared to the $z= [0.4-0.8]$ bins, contains more high-valued WL peaks but fewer low-valued peaks. This behaviour may be avoided by requiring a minimal angular distance between different WL peaks entering the catalog, an effect not dissimilar to fiber collisions.

\subsubsection{Numerical convergence}\label{sec:numerical_convergence}
\begin{figure}
	\includegraphics[width=\columnwidth]{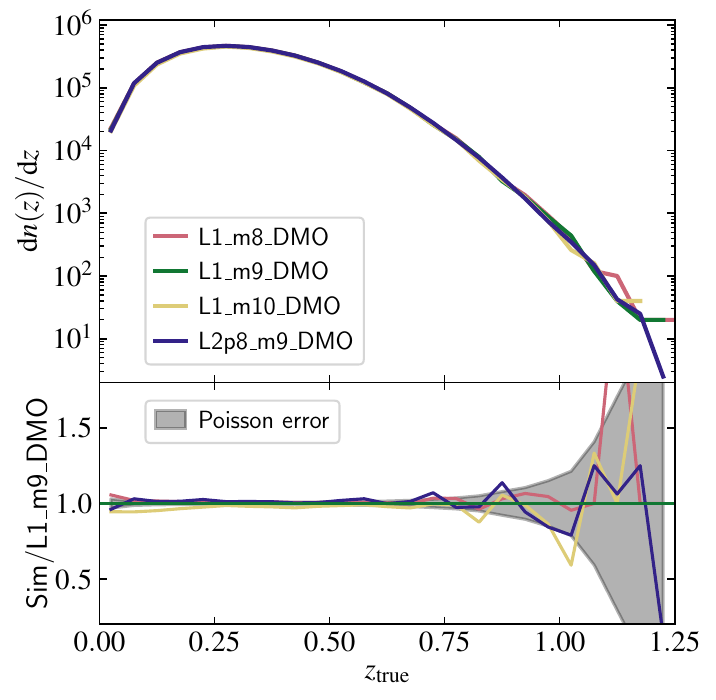}
    \caption{Top: number density of the redshifts of SNR~>~5 WL peaks for L1$\_$m8$\_$DMO (red), L1$\_$m9$\_$DMO (green), L1$\_$m10$\_$DMO (yellow), and the mean of the eight observers in L2p8$\_$m9$\_$DMO (blue). Bottom: Ratio to  L1$\_$m9$\_$DMO. The gray band indicates the Poisson error. The agreement between the variations shows good convergence of the L1$\_$m9 simulations with box size and resolution.}
    \label{fig:numerical_mass_res}
\end{figure}

We now establish the degree of numerical convergence by comparing the redshift distributions as measured for the observers in the simulations with different resolutions and box sizes in the FLAMINGO suite. In Fig.~\ref{fig:numerical_mass_res}, we show the signals measured for the L1$\_$m8$\_$DMO (red), L1$\_$m9$\_$DMO (green), L1$\_$m10$\_$DMO (yellow), and the mean of the eight L2p8$\_$m9$\_$DMO (blue) observers. The lower (m10) and higher (m8) resolution runs have, respectively, 8 times fewer and more particles than m9 in the same volume and with the same initial conditions. We illustrate the redshift distributions as the number density of the redshifts of WL peaks $\mathrm{d} n(z) / \mathrm{d} z$, i.e. the counts per redshift bin over the bin size ($\Delta z = 0.05$). 

The distributions peak at $z \approx 0.3$ but are moderately flat across the $z = 0.2 - 0.4$ range. Whereas the WL kernel already peaks at $z \approx 0.375$, more massive structures continue to form at low $z$, effectively pulling the distribution to lower redshifts than if we had only considered the WL kernel. The bottom panel shows the ratio compared to L1$\_$m9$\_$DMO, with the gray band indicating the Poisson error. The observers in the L1 boxes are located at the same position, and these boxes have the same initial phases. Therefore, comparing these three variations directly shows the degree of numerical convergence with the particle mass and spatial resolution. We find a similar degree of convergence as the number density of WL peaks in \citet{Broxterman2024}. The m10 run shows a systematic offset of a few per cent for $z = [0,0.7]$. As the simulations are initiated with the same phases and based on the convergence of the HMFs of these runs (see fig.~19 of \citet{schaye2023flamingo}), we expect the same massive haloes to exist in the different variations. Because the WL peak sample primarily probes these massive halos (see Fig.~\ref{fig:halo_mass_binning}), we do expect that the SNR~>~5 haloes probed by L1$\_$m9$\_$DMO should still be well resolved in L1$\_$m10$\_$DMO. The small but systematic suppression of the L1$\_$m10$\_$DMO signal shows that the convergence is slightly worse than the convergence of the HMF, which may be a consequence of halo profiles being less resolved and the peak height in the lower resolution simulation not making the SNR~>~5 cut. The convergence between L1$\_$m9 and L1$\_$m8 is much better. There is a 5 per cent difference for the lowest redshift bin, which decreases and remains within 1 per cent up to $z  = 0.7$.

The L2p8$\_$m9 and L1$\_$m9 boxes differ in the realization of the initial conditions. Therefore, they are also impacted by cosmic variance and illustrate the convergence with the volume probed. The convergence with box size is slightly worse than with resolution but still within 3 per cent for z = $0.1-0.8$ and within 1 per cent until for $z = 0.25-0.50$, where the distribution peaks. In general, the difference is similar to the Poisson error for L1$\_$m9.

Whereas Fig.~\ref{fig:numerical_mass_res} shows the numerical convergence for the signal measured in the DMO boxes, we note that the same conclusions apply to the signals measured in the full hydrodynamical simulations.

\subsubsection{Cosmic variance}\label{sec:cosmic_variance}
\begin{figure}
	\includegraphics[width=\columnwidth]{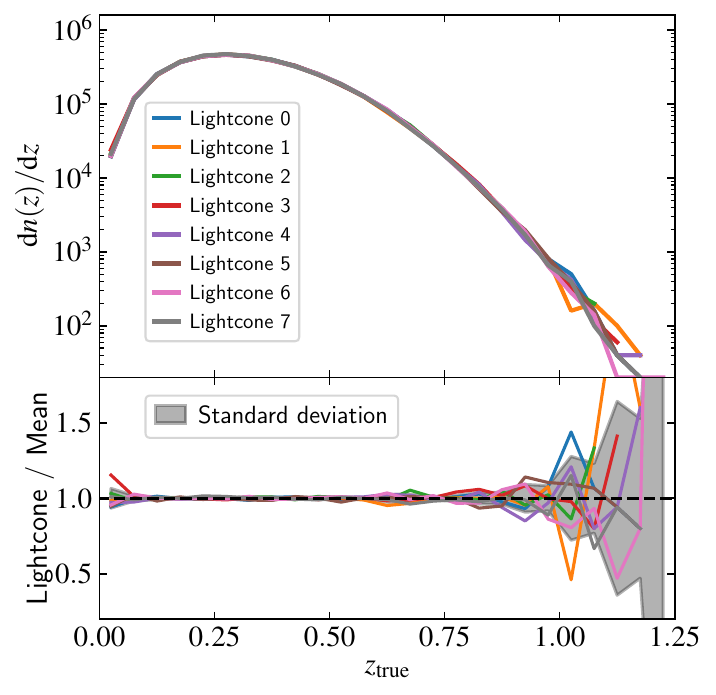}
    \caption{Top: number density of the redshifts of SNR~>~5 WL peaks for the eight virtual observers in L2p8$\_$m9$\_$DMO. Bottom: ratio to the mean redshift distribution viewed by the eight observers. The standard deviation in the bottom panel, shown by the gray-shaded region, estimates the effect of cosmic variance.}
    \label{fig:cosmic_variance}
\end{figure}

Using the eight different observers in the L2p8$\_$m9$\_$DMO box, we test the impact of cosmic variance on our statistics. Again, we show the results for the observers in the DMO run, but we find that the level of difference is the same for the observers in the hydrodynamical simulation. Fig.~\ref{fig:cosmic_variance} shows the redshift distribution of SNR~>~5 WL peaks for the eight virtual observers. The distributions depend on the specific orientation of the individual observer with respect to the most massive haloes in the simulation. The bottom panel shows the ratio of the distribution for each observer to the mean of the eight observers. The degree of cosmic variance is of the same level as the Poisson noise, as shown in the bottom panel of Fig.~\ref{fig:numerical_mass_res}. The effect is less than 1 per cent for $z = 0.1 - 0.6$. 

\subsection{Baryonic and cosmology impact}\label{sec:z_evolution}
\begin{figure*}
	\includegraphics[width=\textwidth]{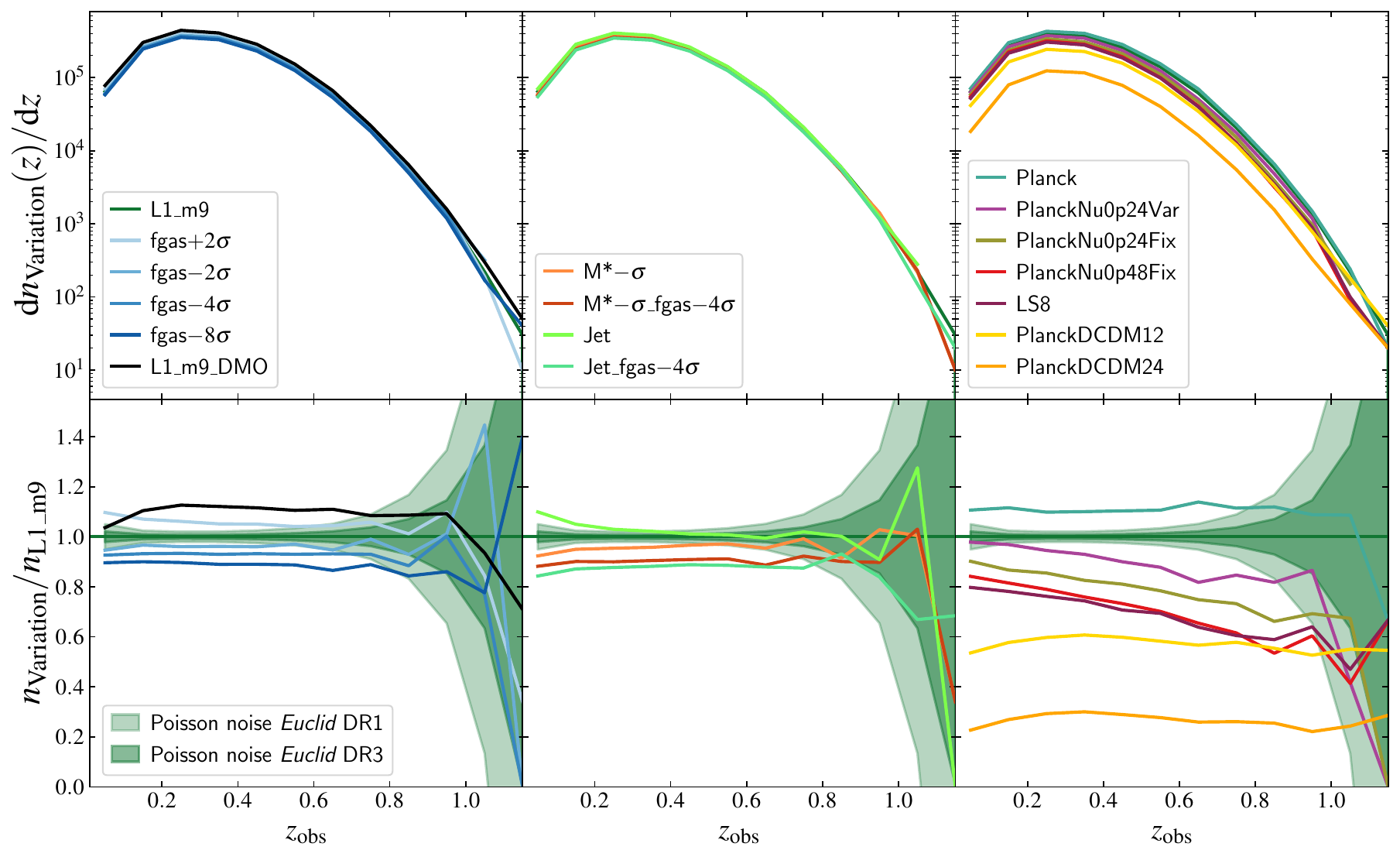}
    \caption{Top: number density of the redshifts of SNR > 5 WL peaks for all L1$\_$m9 FLAMINGO variations. Bottom: Ratio of the redshift distribution of the WL peak number density relative to the prediction for L1$\_$m9 (dark green curve). The left panels includes the variations varying the gas fractions in clusters, with more suppressed gas fractions indicated by an increasingly darker blue colour and the DMO signal in black. The middle panels shows the remaining baryonic variations with a different AGN model (`Jet') or a suppressed galaxy stellar mass function (`M*$-\sigma$'). The right panels shows the cosmology variations, including multiple variations using more massive neutrinos or decaying dark matter (`DCDM'). The shaded green regions in the bottom panels indicate the expected Poisson uncertainty on the \textit{Euclid} DR1 and DR3 measurements. The baryonic variations (left and central panels) show little change in the shape of the redshift distribution compared to the fiducial model, while some of the cosmology variations, particularly the more massive neutrino variations and LS8, show a different shape redshift evolution. The cosmology variations also show a greater overall offset than the baryonic variations. The shape of the redshift distribution of high-SNR WL peaks is insensitive to the intensity of baryonic feedback, but is sensitive to cosmological parameters, making it potentially valuable for cosmological inference.}
    \label{fig:z_evolution_onlysys}
\end{figure*}

Next, we compare the distributions of the virtual observers in the 1~cGpc simulations with different levels of baryonic feedback or cosmologies. We emphasize that all observers in the L1 boxes are located at the same position and that the simulations were initiated with the same initial phases. All SNR~>~5 peaks in the final WL convergence maps are assigned the redshift of the lens plane that contributes most to the WL peak convergence value. At each lens plane, the WL peaks are matched to individual haloes and we assign the redshift of the haloes to the peaks.\footnote{The FLAMINGO halo light-cones for L1$\_$m9$\_$DMO and the decaying dark matter variations are currently unavailable. We, therefore, show the results of these variations using the central redshifts of the shells to which the peaks have been assigned. As we only show relative differences to L1$\_$m9, we also restrict the L1$\_$m9 shell redshifts when comparing to L1$\_$m9$\_$DMO, PlanckDCDM12 and PlanckDCDM24. We confirmed that the ratio to L1$\_$m9 is the same for the simulation variations that do have halo light-cones.} We then adjust the redshifts to account for the expected uncertainty for the massive objects we probe (equation~\ref{eqn:ztrue2zobs}). In Fig.~\ref{fig:z_evolution_onlysys}, we show the ratio of the redshift distributions ($\mathrm{d}n(z)/\mathrm{d}z$) of the WL peaks of all L1 variations (top panels) and their ratio with respect to L1$\_$m9 (bottom panels) as a function of the perturbed observed redshifts $z_{\mathrm{obs}}$. However, we have found that the same conclusions hold when not including any observational redshift uncertainty or catastrophic outliers to the degree of the \textit{Euclid} requirements. We bin the distributions using $\Delta z = 0.1$.

The green shaded region in the bottom panels indicates the Poisson error for the expected number of peaks in \textit{Euclid} data release 1 (DR1, $\sim 2,500$~deg$^2$) and for the final data release (DR3, $\sim 14,000$~deg$^2$). The Poisson errors overestimate the differences between the simulation variations as all L1 variations were initiated with the same initial phases, and we thus expect the simulations to be correlated.

The left and central panels show the predictions for L1$\_$m9$\_$DMO and the baryonic feedback variations. These variations do not show a different redshift evolution from the fiducial L1$\_$m9 model as all curves in the bottom panels are close to horizontal, which even holds for the strongest feedback model, fgas$-8\sigma$. The systematic offsets in amplitude between the curves stem from the fact that the total number of peaks above the SNR cut differs between the simulations. In simulations with stronger feedback, an individual halo that acts as a lens has less mass \citep[e.g.][]{Velliscig2014}. Therefore, the WL convergence value of the peak corresponding to that halo will be lower. As we apply a fixed SNR cut, we expect the total number of peaks to be smaller in simulation with stronger feedback, as was shown in \citet{Broxterman2024} and is reflected in the vertical offsets in Fig.~\ref{fig:z_evolution_onlysys}.

After correcting for the amplitude difference, all deviations from the shape of the fiducial redshift evolution are within a couple of per cent, though the amplitude of the distributions varies by $\sim10$ per cent. The most distinct baryonic variations are L1$\_$m9$\_$DMO and Jet. The DMO prediction shows a minor decrease for $z < 0.2$. However, when quantifying the difference between the distributions in the next Section, we still find the two distributions are statistically similar. The Jet model was calibrated to the same observables as L1$\_$m9, and the only difference is the subgrid implementation of AGN feedback. At $z=0$, Jet shows some enhancement compared to L1$\_$m9, which decreases for higher $z$, and at $z>0.3$, the kinetic and thermal AGN feedback variations give the same result. The comparison indicates some dependence on the simulation's AGN subgrid physics model. Comparing the $z=0$ HMFs of L1$\_$m9 and Jet as shown in Fig.~20 of \citet{schaye2023flamingo}, we observe that the Jet model, compared to L1$\_$m9, has a few per cent more haloes of mass $M_{\mathrm{200c}} = 10^{13}-10^{14}~\mathrm{M_\odot}$. This excess of haloes could explain the overabundance of WL peaks we observe at low $z$.

The lack of differences in the redshift evolution between the baryonic feedback variations is consistent with the interpretation that the redshift distribution of the high-SNR WL peaks is set by the formation time of massive haloes and is insensitive to the strength of the baryonic feedback that manifests itself primarily only once a halo has formed and has accreted sufficient mass. The strength of the feedback does not impact the time at which a halo has formed but only whether it creates a WL signal significant enough to be included in the SNR > 5 sample, as reflected in the amplitude differences as shown in Figure~\ref{fig:z_evolution_onlysys}. This reasoning still holds if baryonic feedback is redshift dependent or if haloes of the same mass experience different baryonic feedback at different redshifts, provided the WL peaks correspond to single massive objects, as we have shown is the case, and the baryonic evolution is independent of cosmology, which is to first order the case for WL peaks measured in the limited number of FLAMINGO variations at different cosmologies. As the shape of the L1$\_$m9$\_$DMO redshift evolution is consistent with that for the hydrodynamical simulations assuming the same cosmology, the redshift evolution of the WL peak sample studied in this work may be modeled with $N$-body simulations only. 

Instead, if we focus on the cosmology variations in the right panels, we see greater differences in amplitude and evolution. We distinguish two subgroups in these panels. The first group, consisting of the neutrino mass variations and LS8, shows a different redshift evolution compared to the runs using the fiducial cosmology. Neutrinos act as a form of hot DM, thereby suppressing structure formation \citep[e.g.][]{Lesgourgues2006neutrinos}{}{}. This is reflected in the lower value of $\sigma_\mathrm{8}$, as shown in the next to last column in Table~\ref{tab:simulations}. LS8, which has a lower value of $\sigma_8$ by design, and the higher neutrino mass variations have a lower value of $\sigma_\mathrm{8}$ than the fiducial cosmology. Consequently, there are fewer massive haloes in these simulations. This difference in the number of massive clusters in these simulations, compared to L1$\_$m9, increases with redshift. This is also reflected in the redshift distributions, as those belonging to the more massive neutrino and LS8 variations are increasingly more suppressed at higher $z$. 

The second group, consisting of the Planck and decaying dark matter models, shows no strong difference in the redshift evolution compared to L1$\_$m9. The \textit{Planck} cosmology has a 3.3 per cent higher value of $\Omega_{\mathrm{m}}$ and only a 0.6 per cent lower value of $\sigma_8$. Considering the HMF, larger values of $\Omega_{\mathrm{m}}$ increase the overall amplitude of the HMF, whereas a lower $\sigma_8$ moves the exponential drop to lower halo masses \citep[e.g.][]{Xhakaj2023hmf}{}{}. Assuming all peaks emanate from a single halo, we expect the  Planck and L1$\_$m9 (i.e. DES Y3 ‘3 × 2pt + All Ext.’) to differ primarily in the total number of haloes as the difference in $\Omega_{\mathrm{m}}$ is a factor 5 greater than the difference in $\sigma_8$. This is reflected in the redshift distributions, as the Planck model has more peaks overall, but there is no apparent difference in the redshift evolution. However, more variations should be compared to investigate the impact of different cosmological parameters on the WL peak redshift evolution.

In the DCDM variations, each dark matter particle, independent of its velocity or local density, loses part of its mass as set by the decay rate indicated in the final column of Table~\ref{tab:simulations}. The half-life of these models is relatively short compared to values that have already been ruled out by geometrical constraints from the BAO \citep[e.g.][]{Aubourgh2015}, but the variations can still be used to understand the impact of decaying dark matter qualitatively. As the DM decays, the total matter density decreases, as indicated in the 9th column of Table~\ref{tab:simulations}, and the growth of structure formation slows down \citep[e.g.][]{Aoyama2014,Enqvist2015}. This is reflected in Fig.~\ref{fig:z_evolution_onlysys} as the suppression of the PlanckDCDM12 and 24 models is, respectively, a factor 2 and 3 greater than for the other variations. The decaying dark matter models are variations on the Planck model, so we expect them to converge back to Planck at high redshift. Whereas visually the DCDM models do not show large differences from L1$\_$m9, in the next Section, we will quantify the difference in redshift evolution; we find that the DCDM variations are still discrepant to the degree that we expect \textit{Euclid} to be able to discriminate between them based on the shape of the redshift distribution of SNR~>~5 peaks. The mass loss in the DCDM becomes increasingly more apparent at low redshift. This is reflected in the DCDM signals as the suppression increases for $z < 0.4$, which can be interpreted as a larger fraction of low$-z$ lenses that lose enough mass to drop below the SNR threshold.

The comparison illustrates that the redshift evolution of WL peaks can constrain parts of the cosmological parameter space; as we have illustrated, it is sensitive to the neutrino mass. However, it also shows that changes in cosmology may result in degenerate signatures when using only WL peaks. For example, the most massive neutrinos run and LS8 are practically indistinguishable. 

In \citet{Broxterman2024}, we showed that even though the impact of cosmology on the number density of the WL peaks is greater than that of realistic baryonic feedback, the two signatures are degenerate. In this work, we have shown that changes in cosmology impact the redshift evolution of the number density of high-SNR WL peaks more than baryonic feedback is expected to do. Combining the redshift and WL convergence distributions can, therefore, not only help constrain cosmology, but also simultaneously calibrate baryonic feedback. The evolution of the redshift distribution of the WL peaks will yield the cosmology, while the number density of WL convergence peaks is sensitive to the baryonic feedback strength. However, as indicated by the Planck variation, not all cosmologies change the redshift distribution of the peaks. Baryonic feedback and changes in cosmology are thus still degenerate in certain parts of the parameter space. From the variations we explored, this seems to be the case for $\Omega_\mathrm{m}$ and baryonic feedback strength leading to lower gas fractions in low-$z$ clusters. However, more variations that jointly vary baryonic physics and cosmology should be compared in order to study this in more detail. 

Ultimately, baryonic physics could be calibrated jointly while constraining cosmology by combining different WL statistics \citep[see e.g.][]{Semboloni2013}, that are sensitive in different ways to cosmology and baryonic physics. This may be extended by including (the cross-correlation with) additional probes to inform on the strength of baryonic, such as SZ or X-ray \citep[see e.g.][]{McCharty2023,McCarthy2024,Bigwood2024}. In this section, we have argued that WL peaks will be useful in furthering this goal.

\subsection{Cosmology with the redshift of WL peaks}\label{sec:ks_test_sec}
Next, we quantify the differences between the simulation variations to L1$\_$m9 to determine the feasibility of using WL peaks to constrain cosmology with Stage IV WL surveys. As we are interested in the shape of the distribution, as opposed to its amplitude, we use the Kolmogorov–Smirnov test. The non-parametric test evaluates the null hypothesis that two distributions are equal by comparing their cumulative density functions. For each of the L1 variations, we randomly draw the number of peaks corresponding to our \textit{Euclid} DR3 estimate (i.e. corresponding to $14,000$~deg$^{2}$ instead of the full light-cone) and compare this to the L1$\_$m9 distribution, which yields a $p$-value describing the hypothesis of the distributions being identical. We randomly draw a WL peak sample 1000 times and report the mean of the $p$-values for the WL convergence and redshift number densities in the second and third columns of Table~\ref{tab:number_of_sigma}, respectively. Entries with a greater value are more consistent with the fiducial simulation. In bold, we highlight all entries with a $p$-value lower than 0.1 to illustrate variations whose distribution differs most from L1$\_$m9. The differences in the $\kappa$ distributions are generally greater than those in the redshift distributions. Similarly, the differences induced by the changes in cosmology are greater than those between the baryonic feedback variations.

\begin{table}
\caption{$p$-value of the KS test of the null hypothesis that the shape of the number density as a function of WL convergence ($p_\kappa$) or redshift ($p_z$) of the SNR~>~5 WL peak distribution measured in a simulation variation is identical to L1$\_$m9. A high $p$-value indicates that the two distributions are statistically indistinguishable. Values with a $p$-value lower than 0.1 are indicated in bold. The shape of the WL convergence number densities differs between cosmology and baryonic variations. The shape of the redshift distribution depends primarily on cosmology and, to a much smaller degree, on baryonic feedback.}
\label{tab:number_of_sigma}
\centering
\begin{tabular}{lll}
\hline
Identifier     & $p_\kappa $   & $p_z$ \\
\hline
L1$\_$m9$\_$DMO & 0.43  & 0.18 \\
fgas$+2\sigma$ & 0.38  & 0.16 \\
fgas$-2\sigma$ & 0.47  & 0.64 \\
fgas$-4\sigma$ & 0.16  & 0.67 \\
fgas$-8\sigma$ & $\boldsymbol{7.9 \times 10^{-3}}$  & 0.18 \\
M*$-\sigma$    & 0.38  & 0.33 \\
M*$-\sigma\_$fgas$-4\sigma$     &  $\boldsymbol{1.8 \times 10^{-2}}$  & 0.62 \\
Jet            &  0.26 & $\boldsymbol{6.6 \times 10^{-3}}$ \\
Jet$\_$fgas$-4\sigma$      & $\boldsymbol{5.4 \times 10^{-3}}$   & 0.45 \\
Planck         &  $\boldsymbol{3.4 \times 10^{-2}}$ & 0.56 \\
PlanckNu0p24Var & $\boldsymbol{7.2 \times 10^{-3}}$  & $\boldsymbol{< 10^{-5}}$ \\
PlanckNu0p24Fix &  $\boldsymbol{< 10^{-5}}$  &  $\boldsymbol{< 10^{-5}}$\\
PlanckNu0p48Fix &  $\boldsymbol{< 10^{-5}}$  &  $\boldsymbol{< 10^{-5}}$ \\
PlanckDCDM12   & $\boldsymbol{< 10^{-5}}$  & $\boldsymbol{3.7 \times 10^{-2}}$ \\
PlanckDCDM24   &  $\boldsymbol{< 10^{-5}}$  & $\boldsymbol{1.2 \times 10^{-3}}$ \\
LS8            &  $\boldsymbol{< 10^{-5}}$ &   $\boldsymbol{< 10^{-5}}$ \\
\hline
\end{tabular}
\end{table}

Focusing first on the difference in the $\kappa$ distributions, we expect all cosmology variations to result in number density distributions that are discrepant with L1$\_$m9 to the degree that should be measurable with the \textit{Euclid} DR3 sample. At the same time, we see that some of the baryonic variations, specifically the simulations calibrated to $f_\mathrm{gas}-8\sigma$, M*$-\sigma\_$fgas$-4\sigma$ and Jet$\_$fgas$-4\sigma$, also result in similar differences. This means the baryonic feedback should be understood and modeled accurately before the $\kappa$ number density of WL peaks can be exploited to constrain cosmology.

In contrast, based on the bold-valued entries for the $p_z$-values, we generally find that the baryonic feedback variations are consistent with L1$\_$m9, whereas the cosmology variations show great differences as all but Planck and the DCDM variations have $p$-values lower than $10^{-5}$. Only one baryonic feedback model, the Jet model, is found to have a redshift distribution with a shape statistically different from L1$\_$m9. This is not the case for the Jet model with stronger feedback, Jet$\_$fgas$-4\sigma$. The origin of this difference is unclear, but may indicate some dependence on the simulation's AGN subgrid model.

If we carry out a similar test to compare fgas$-8\sigma$ to PlanckNu0p24Var we find that the corresponding $p$-values are $p_\kappa = 0.48 $ and $p_z = 6.5 \times 10^{-5}$, i.e. the shape of their distributions in WL convergence is indistinguishable while their redshift distributions are different. As these variations differ in terms of both cosmology and baryonic feedback, the comparison highlights the degeneracy between cosmology and baryonic feedback in the number density in WL convergence, which may be broken by including the redshift information.

To conclude, the $p_z$ values confirm the result that the redshift distribution of WL peaks depends primarily on cosmology and, to a much smaller degree, on baryonic feedback strength. Therefore, the redshift distribution of high-valued SNR peaks provides an excellent probe to help constrain cosmology complementary to the commonly used peak height distribution.

\section{Summary and Conclusions}\label{sec:sum_and_concl}
In this paper, we have studied the cosmological potential of the redshift distribution of the expected measurement of high-$\kappa$ (SNR~>~5) WL peaks for a Stage IV WL survey using the FLAMINGO cosmological hydrodynamical simulation suite. We measured the contribution to the WL convergence value along the line of sight for WL peaks that were selected from integrated maps, and we assigned the redshift of the most contributing overdensity to each peak. At the peak redshift, we match peaks to haloes to determine the origin of the WL peaks. We find that high-SNR peaks (SNR~>~5) originate primarily from single massive haloes with $M_{\mathrm{200c}} > 10^{14}~\mathrm{M_\odot}$ (Fig.~\ref{fig:purity}) and that more massive haloes preferentially cause higher$-\kappa$ peaks (Fig.~\ref{fig:halo_mass_binning}). We estimate Stage IV WL surveys will detect $5.5\times10^{4}$ ($6.8\times10^{3}$) SNR~>~5 (8) WL peaks with a purity, i.e. fraction belonging to a single massive halo along the line of sight, of 76 (96) per cent. For haloes with $M_{\mathrm{200c}} > 10^{14.5}~\mathrm{M_\odot}$, we find a completeness, i.e. the percentage of haloes that is matched to a peak in our SNR~>~5 peak sample, of 93 (59) per cent up to $z=0.5$ ($1$) (Fig.~\ref{fig:completeness_with_z}). The purity and completeness are competitive with state-of-the-art X-ray and SZ cluster abundance inferences. The paper has highlighted the possibility of using the WL peaks measured by upcoming Stage IV WL surveys. In the future, the distribution of the number density of WL peaks in WL convergence and redshift may be studied as a function of cosmology, as is done commonly for the HMF. 

By matching each WL peak to individual haloes and by assigning a redshift, we have studied the impact of baryonic physics and cosmology on the redshift evolution of the number density of WL peaks. We have shown that, within the FLAMINGO simulation suite, the changes in baryonic feedback strength, as parameterized by the gas fraction in clusters and/or the $z=0$ galaxy stellar mass function, do not have a substantial impact on the redshift evolution of the sample of high-SNR WL peaks that will be measured by Stage IV WL surveys (Fig.~\ref{fig:z_evolution_onlysys}). This includes models where the gas fractions in low$-z$ clustered are lowered by 8$\sigma$ compared to the fiducial model and a DMO+$\nu$ simulation without baryons. 

In contrast, changes in cosmology can have a strong impact on the redshift distribution of high-SNR WL peaks. From the variations we have explored, we find the redshift distribution is particularly sensitive to the neutrino mass, which impacts structure formation and the value of $\sigma_8$. However, more variations that jointly vary cosmology and baryonic physics should be carried out to further quantify which regions of cosmology and baryonic physics can be effectively constrained using the redshift distribution of WL peaks. Combining the redshift distribution of WL peaks with their number density as a function of the WL convergence can help simultaneously calibrate baryonic feedback and constrain cosmology, as the two statistics show different dependencies on baryonic physics. The constraining power can be increased by including other higher-order WL statistics or by including (cross-correlations with) other observables, such as SZ or X-ray.

\section*{Acknowledgements}
RK acknowledges support by research programme Athena 184.034.002 from the Dutch Research Council (NWO). VJFM acknowledges support by NWO through the Dark Universe Science Collaboration (OCENW.XL21.XL21.025). This work used the DiRAC@Durham facility managed by the Institute for Computational Cosmology on behalf of the STFC DiRAC HPC Facility (\url{www.dirac.ac.uk}). The equipment was funded by BEIS capital funding via STFC capital grants ST/K00042X/1, ST/P002293/1, ST/R002371/1 and ST/S002502/1, Durham University and STFC operations grant ST/R000832/1. DiRAC is part of the National e-Infrastructure. We have relied on the numerical packages \textsc{matplotlib} \citep[][]{Hunter2007}, \textsc{healpy} \citep[][]{Zonca2019Healpy}, \textsc{numpy} \citep[][]{Harris2020}, \textsc{swiftsimio} \citep[][]{borrow2020}, and \textsc{astropy} \citep[][]{Astropy2022} for the analysis and plotting. 

\section*{Data Availability}
The data supporting the figures in this article are available upon reasonable request to the corresponding author.



\bibliographystyle{mnras}
\bibliography{example} 




\appendix

\section{Smoothing} \label{app:smoothing}

\begin{figure*}
	\includegraphics[width=\textwidth]{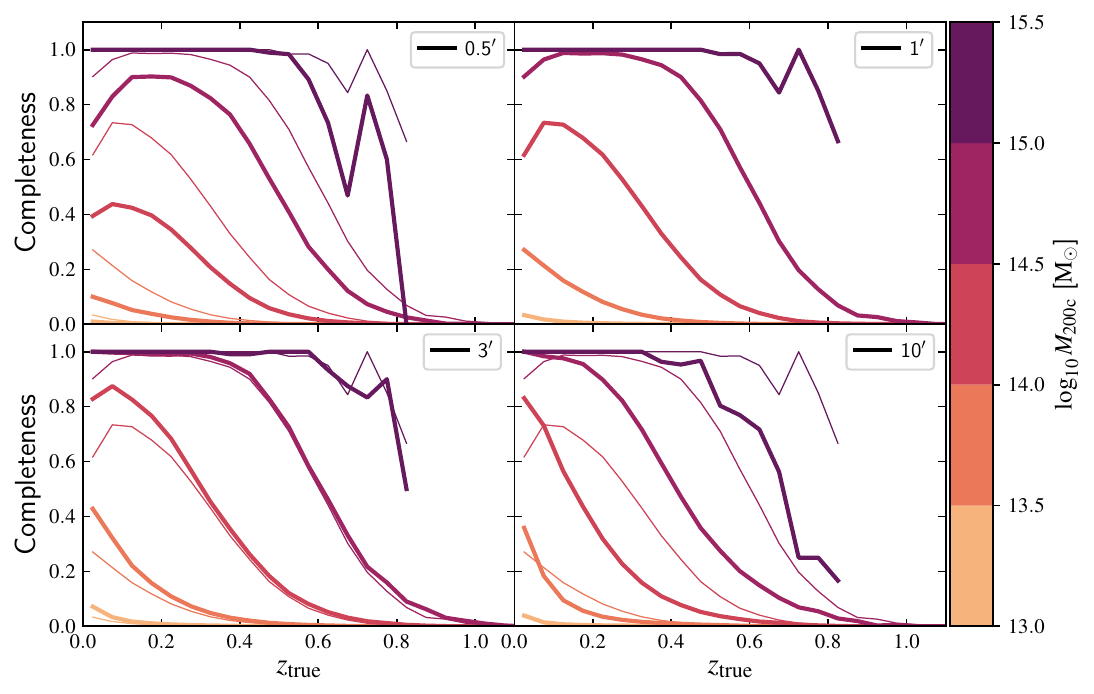}
    \caption{Completeness, i.e. the fraction of haloes within a halo mass bin matched to an SNR~>~5 WL peak for the four different smoothing scales, as indicated in the upper right of each panel. The results for 1 arcmin smoothing are repeated in all panels as thin curves. For small smoothing scales, some massive haloes at low redshift that span an angular size larger than is smoothed over will be missed as the smoothing scale is too small to achieve the optimal SNR. At the same time, increasing the smoothing scale will result in lower completeness at high redshift as those objects subtend smaller sizes, and increasing the smoothing will decrease the SNR of those objects. Combining smoothing scales may thus improve the amount of information that is extracted.}
    \label{fig:smoothing_completeness}
\end{figure*}

In this Appendix, we study the impact of the adopted smoothing scale on the halo peak matching. The fiducial smoothing scale adopted in this work is 1 arcmin. In Fig.~\ref{fig:smoothing_completeness}, we show the results of the completeness of the halo peak matching for three additional smoothing scales: $0.5~\arcmin$ (upper left), $1~\arcmin$ (upper right), $3~\arcmin$ (lower left), and $10~\arcmin$ (lower right). The curves corresponding to $1~\arcmin$ smoothing are repeated as thin lines in each panel.

For each smoothing scale, the figure shows the completeness, i.e. the fraction of haloes matched to a peak (with SNR~>~5) within a halo mass bin. As already mentioned in Section~\ref{sec:completeness}, the small drop in completeness at low redshift for the $M_{\mathrm{200c}} = 10^{14}-10^{15}~\mathrm{M_\odot}$ mass bins is the result of the smoothing scale being too small to achieve the optimal SNR for these objects. At $z = 0.1$, the smoothing scale of $1~\arcmin$ corresponds to a physical scale of $\approx110$~kpc, which is an order of magnitude smaller than the viral radius of $M_{\mathrm{200c}} \ge 10^{14}~\mathrm{M_\odot}$ haloes.

This smoothing scale is, therefore, too small to achieve the optimal SNR for these haloes. This is illustrated by the fact that the completeness of the same halo mass bins is larger when $3~\arcmin$ smoothing has been applied. The most massive haloes ($M_{\mathrm{200c}} \ge 10^{15}~\mathrm{M_\odot}$) do not suffer from this problem, as they are so massive that independent of the smoothing scale, they will cause a large enough WL peak to be detected. A larger smoothing scale than the fiducial $1~\arcmin$ will allow the detection of more low-$z$ objects as this scale corresponds better to the angular size of the objects we are probing. However, simultaneously, the completeness at higher $z$ drops for larger smoothing scales as the objects have a smaller angular scale at higher redshifts.  

Ideally, the smoothing scale should be matched to the redshift of an object such that an appropriate scale can be chosen that smooths over the entire object but not over a larger area than necessary, as having either a too-small or too-large smoothing scale will decrease the SNR of the object. As the redshift of an object is not known beforehand, it is not clear what the optimal smoothing scale is to detect an object until it has already been detected and identified with an optical counterpart, and the choice of smoothing scale is a trade-off between detecting more objects at higher redshift against lower completeness at lower redshift. We chose a $1~\arcmin$ smoothing scale as the fiducial scale to keep the results consistent with \citet{Broxterman2024}. However, we already noted that multiple smoothing scales may be combined to probe information from different scales.

\section{Noise peak distribution}\label{app:noise_peaks}

\begin{figure}
	\includegraphics[width=\columnwidth]{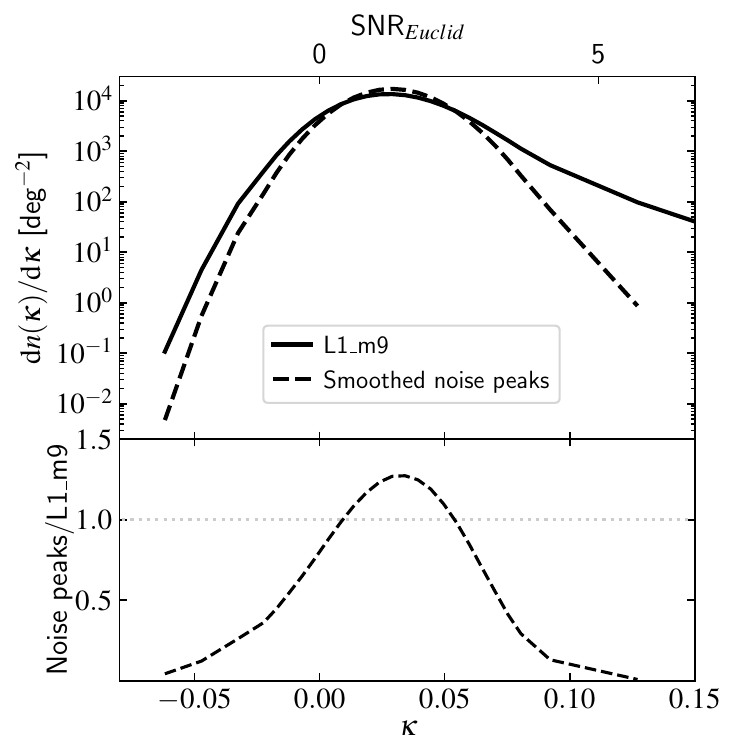}
    \caption{Top: number density of WL peaks for the WL convergence map that includes noise and smoothing (solid curve) and of only the smoothed noise (dashed curve).  Bottom: ratio of the smoothed noise peak distribution to L1$\_$m9. The smoothed noise peak distribution dominates the signal for $ 0 \lesssim \kappa \lesssim 0.07$, and lower and higher valued peaks generally correspond to physical objects.}
    \label{fig:noise_plot}
\end{figure}

In this Appendix, we compare the number density of WL peaks for our \textit{Euclid}-like sample to a distribution generated only by noise. The noise field is generated as described in Section~\ref{sec:peaks}. In the regular procedure, we add the noise to the WL convergence field and then smooth the map. The number density of WL peaks selected from this map is indicated by the solid curve in Fig.~\ref{fig:noise_plot}. Now, we also show the peak distribution that is obtained if we directly determine the peaks from the smoothed noise map (`Smoothed noise peaks'), which do not contain any cosmological information. The Smoothed noise peak distribution is indicated by the dashed curve. Whereas the noise is a normal distribution centred on $\kappa = 0$, the Smoothed noise peak distribution, consisting of the local maxima of the smoothed noise map, has a positive mean and an asymmetric shape. 

In Fig.~\ref{fig:purity}, we illustrated that the Noise peak distribution peaks at $\kappa \approx 0.04$. Fig.~\ref{fig:noise_plot} illustrates that this is also the WL convergence value at which the distribution from the applied noise field peaks. As expected, for larger $\kappa$ values, the distribution of Smoothed noise peak decreases, and the majority of the L1$\_$m9 peaks are no longer caused by the noise, as also already shown in Fig.~\ref{fig:purity}. Similarly, the lowest valued WL peaks, with $\kappa \lessapprox$ 0, are also increasingly less likely to be caused by the random noise. These peaks, therefore, correspond to the local maxima in void regions in the simulations. 

\section{Halo-peak matching characteristics}\label{app:extra_field}

\begin{figure*}
	\includegraphics[width=\textwidth]{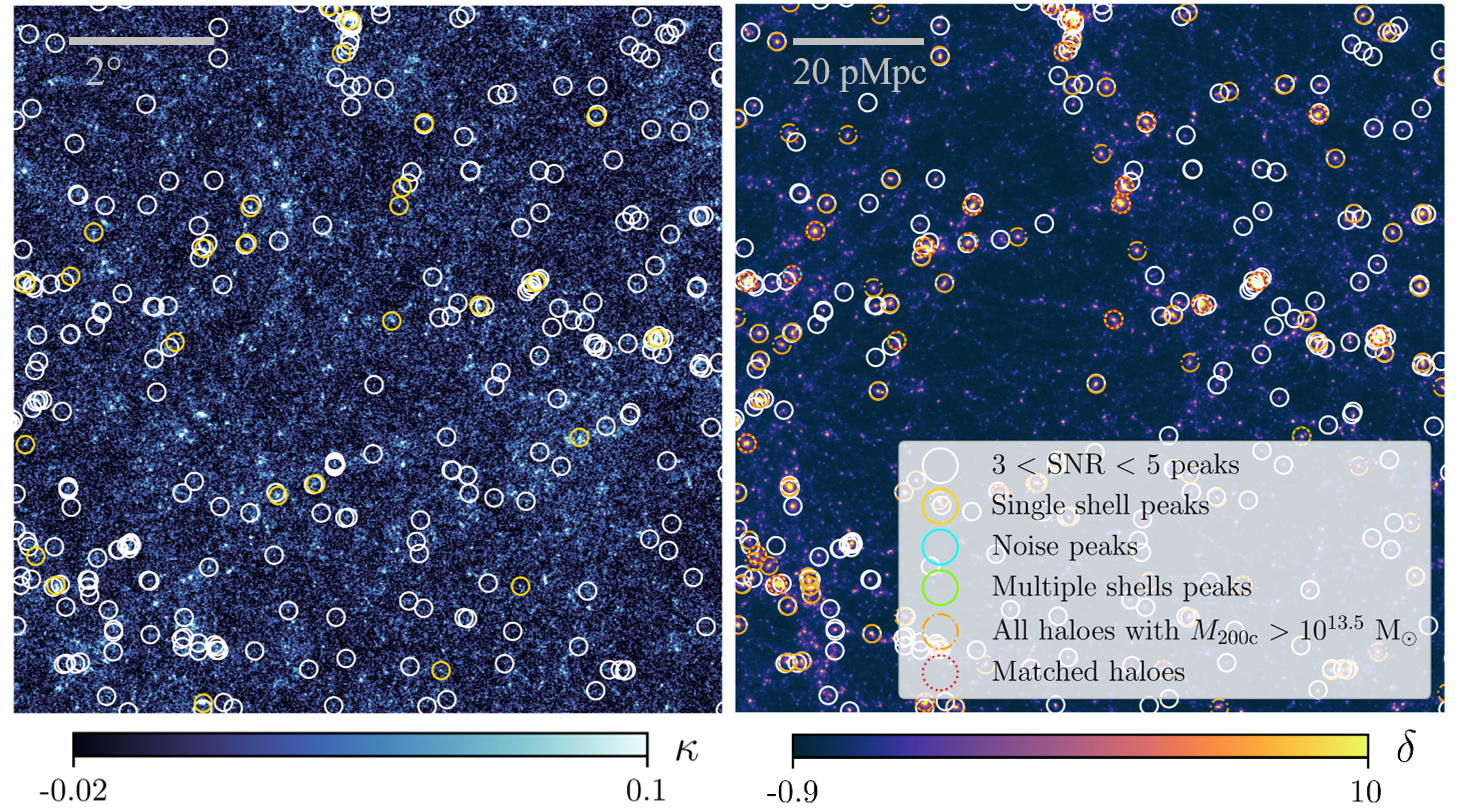}
    \caption{Same as Fig.~\ref{fig:gnomview_plot} and now also showing the different peak categories to which the peaks belong. The Single shell peaks (solid yellow), Noise peaks (solid cyan), Multiple shells peaks (solid green), peaks assigned to this shell with SNR~=~$3-5$ (solid white), all $\log_{\mathrm{10}}M_{\mathrm{200c}}[\mathrm{M_\odot}] > 13.5$ haloes (dashed-dotted orange) and haloes that have been matched to an SNR~>~5 peak (dotted red). Most Noise and Multiple shells peaks are still visibly well matched to a halo at their position. The apparent overdensities that do not generate an SNR~>~5 peak produce peaks with an SNR value lower than this cut.}
    \label{fig:gnomview_plot_extra}
\end{figure*}

In this Appendix, we study the characteristics of the peak-halo matching algorithm in more detail. In Fig.~\ref{fig:gnomview_plot_extra}, we show the same fields as in Fig.~\ref{fig:gnomview_plot}, but we split the peaks into the three peak categories we considered in this work. The solid yellow, cyan, and green circles indicate the Single shell, Noise, and Multiple shells peaks, respectively. We also show all $M_{\mathrm{200c}} > 10^{13.5}~\mathrm{M_\odot}$ haloes from the FLAMINGO halo light-cone as dashed-dotted white circles and the matched haloes as dotted red circles. 

Fig.~\ref{fig:gnomview_plot_extra} shows that Noise and Multiple shells peaks may still correspond to massive overdense objects. For example, a Noise peak is indicated by the solid cyan circle in the middle left of the figure. Similarly, the solid green circle directly above the legend shows that this Multiple shells peak can also be assigned to a halo at the same angular position. When comparing the redshift distribution of the peaks, we have also assigned a redshift to the Noise peaks and Multiple shells peaks. In any observational inference, we will not know which peaks are Noise peaks or Multiple shells peaks, but can only assign a redshift based on the galaxies at the position of the peak. The comparison here shows that this is a valid approach, as there are still haloes assigned that contribute significantly. 

In orange, we show the WL peaks assigned to this shell with an SNR of $3-5$. These are lower values than the standard range we consider in this work. The figure shows that most of these peaks still clearly correspond to an overdense structure in the field, as indicated by the white dashed-dotted circles. Lowering the SNR threshold would allow us to include these matches in samples we consider and possibly get tighter constraints, but this comes at the cost of decreasing the purity of the sample, as shown in Fig.~\ref{fig:purity}. As a result, the interpretation will be more difficult, as a superposition of objects or noise will cause more WL peaks, and interpreting them each as arising from a single halo along the line of sight will no longer be justified. This is also clear from the fact that some of these peaks do not correspond to a massive overdense structure.  However, we stress that our chosen SNR cut-off may not be optimal. Additional work should be done to explore the optimal SNR cut-off value to balance the trade-off between cosmological information and purity.

\bsp	
\label{lastpage}
\end{document}